\documentclass[a4paper,onecolumn,11pt,accepted=2017-05-09]{quantumarticle}
\pdfoutput=1
\usepackage[utf8]{inputenc}
\usepackage[english]{babel}
\usepackage[T1]{fontenc}
\usepackage{amsmath}
\usepackage{amsfonts}
\usepackage{hyperref}

\usepackage{tikz}
\usepackage{lipsum}

\begin{document}

\title{Complete security analysis of {quantum key distribution} based on unified model of sequential discrimination strategy}

\author{Min Namkung$^{2,}$}
\affiliation{Center for Quantum Information, Korea Institute of Science and Technology (KIST), Republic of Korea}
\author{Younghun Kwon}
\affiliation{Department of Applied Physics, Hanyang University (ERICA)}
\email{yyhkwon@hanyang.ac.kr}
\maketitle

\begin{abstract}
  The quantum key distribution for multiparty is one of the essential subjects of study. Especially,  without using entangled states, performing the quantum key distribution for multiparty is a critical area of research. For this purpose, sequential discrimination, which provides multiparty quantum communication and quantum key distribution for {multiple receivers}, has recently been introduced. However, since there is a possibility of eavesdropping on the measurement result of a receiver by an intruder using quantum entanglement, a security analysis for {quantum key distribution} should be performed. {However,} no one has provided the security analysis for {quantum key distribution in view of the sequential scheme} yet. In this work, by proposing a unified model of sequential discrimination including an eavesdropper, we provide the security analysis of {quantum key distribution based on the unified model of sequential discrimination strategy.} In this model, the success probability of eavesdropping and the secret key rate can be used as a figure of merit.  Then, we obtain a non-zero secret key rate between the sender and receiver, which implies that the sender and receiver can share a secret key despite eavesdropping.
Further, we propose a realistic quantum optical experiment for the proposed model. We observe that the secret key between the sender and receiver can be non-zero, even with imperfections. As opposed to common belief, we further observe that the success probability of eavesdropping is smaller in the case of colored noise than in the case of white noise.
\end{abstract}

\section{Introduction}
 Quantum physics restricts perfect quantum state discrimination(QSD), which contradicts the argument of classical physics \cite{a.chefles,s.m.barnett,j.a.bergou,j.bae}. This fact takes a major role in quantum information processing. According to the optimal strategy of QSD required in terms of the figure of merit, there exist well-known strategies such as minimum error discrimination \cite{c.w.helstrom,a.s.holevo,h.p.yuen,c.l.chow,u.herzog,j.bae3,d.ha,d.ha2,d.ha3,d.ha6}, unambiguous discrimination \cite{i.d.ivanovic,d.dieks,a.peres,g.jaeger,t.rudolph,u.herzog2,s.pang,j.a.bergou2,d.ha4}, maximal confidence \cite{s.croke}, and a fixed rate of inconclusive results \cite{a.chefles2,c.w.zhang,j.fiurasek,y.c.eldar,u.herzog3,e.bagan,k.nakahira,u.herzog4,d.ha5}, which can be applied to two-party quantum communication.

\indent There can be many receivers in quantum communication, and the strategy of QSD between two parties needs to be extended to multiple parties. In 2013, Bergou et al.\cite{j.a.bergou3} proposed sequential discrimination in which many parties can participate as receivers.  Sequential discrimination is process in which the post-measurement state of a receiver is passed to the next receiver. The fact that the probability that every receiver can succeed in discriminating the given quantum state is nonzero implies that every receiver can obtain the information of the quantum state of the sender, from the post-measurement state of the preceding receiver \cite{p.rapcan,c.-q.pang,z.-h.zhang,m.hillery,m.namkung,m.namkung2}. It was shown that sequential discrimination can provide multiparty B92 protocol \cite{m.namkung3}, which was implemented using quantum optical experiment \cite{m.a.solis-prosser,m.namkung4}.

When sequential discrimination is performed, one
 can assume that an eavesdropper may exist. 
Suppose that Alice and Bob performs quantum communication through the B92 protocol and Eve tries to eavesdrop. The eavesdropper can have two ways for eavesdropping. The first situation is the case where Eve tries to eavesdrop on Alice's quantum state, which was analyzed in  \cite{m.namkung2}. The second situation is where Eve tries to eavesdrop on the result of Bob. \\
\indent Even though the second situation is a major threat to secure communication, the security analysis to this case has not been done yet. {Therefore, in this paper, we focus on the second case, in which an intruder tries to eavesdrop on the result of a receiver, and provide a systematic security analysis from a unified model of sequential discrimination including an eavesdropper.} In this proposed model, the success probability of eavesdropping and the secret key rate \cite{i.csiszar} can be considered as a figure of merit \textit{for the security analysis}. Specifically, the figure of merit for Eve is the success probability of eavesdropping, but the figure of merit for Alice and Bob is the secret key rate. Our study shows that although Eve performs an optimal measurement for the success probability of eavesdropping, the secret key rate between Alice and Bob is not zero.\\
\indent In addition, we propose a quantum optical experiment that implements a new sequential discrimination method composed of Alice-Eve-Bob. The quantum optical experiment consists of a linear optical system similar to a Sagnac interferometer \cite{m.a.solis-prosser,f.a.torres-ruiz}. The experimental setup can achieve an optimal success probability of eavesdropping. Further, we provide the success probability of eavesdropping and the secret key rate, considering the imperfections that can occur in the source, channel, and detector. White noise and colored noise are considered imperfections of the source \cite{a.cabello}. The dark count rate and detection efficiency are considered imperfections of the detector \cite{g.cariolaro}. 

In this paper, we consider security analysis of the B92 protocol in view of the sequential discrimination scheme. That is because the security analysis can be performed with the simple mathematical structure of the unambiguous discrimination in this scheme \cite{j.a.bergou2,m.namkung2}. We emphasize that our methodology based on the sequential discrimination can be applied to the various kinds of quantum communication \cite{i.a.burenkov} as well as quantum key distribution \cite{c.h.bennett2} designed in prepare-and-measure way. Moreover, our scheme can be applied to quantum communication or key distribution task utilizing the continuous variable quantum systems \cite{g.cariolaro,a.serafini}. We further emphasize that our research propose a novel theoretical way to unify the secure quantum communication tasks in terms of the quantum state discrimination.

\begin{figure*}[t]
\centerline{\includegraphics[width=14cm]{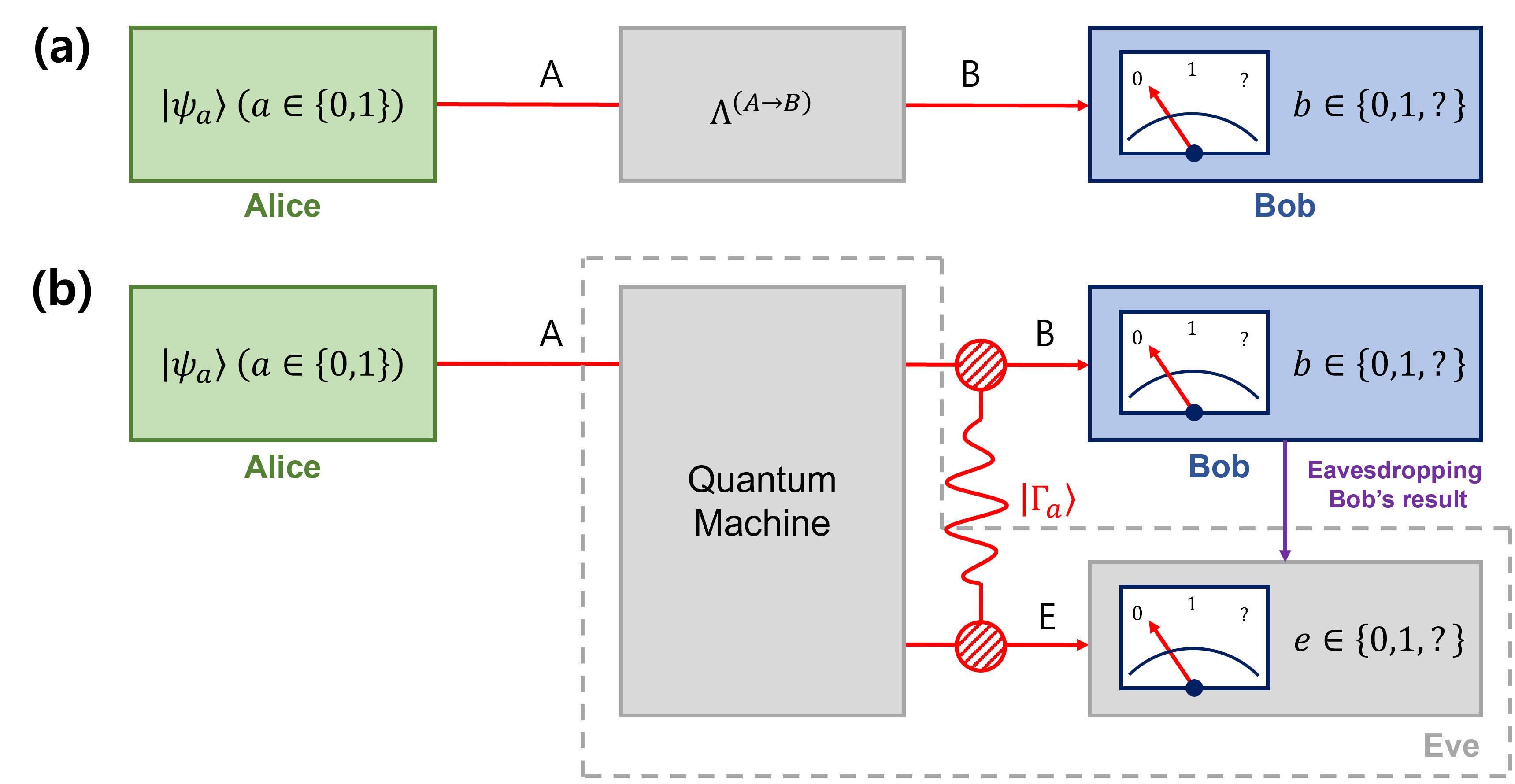}}
\caption{{Eve's scheme for eavesdropping Bob's measurement result. If Eve is unnoticed by Alice and Bob, as illustrated as (a), then the quantum channel between Alice and Bob is described as a depolarizing channel $\Lambda^{(A\rightarrow B)}$. In this scheme, Eve uses a quantum machine that deterministically transforms Alice's state $|\psi_a\rangle$ to a composite system $|\Gamma_a\rangle$ such that $\mathrm{Tr}_E\left(|\Gamma_a\rangle\langle\Gamma_a|\right)=\Lambda^{(A\rightarrow B)}(|\psi_a\rangle\langle\psi_a|)$. Then, she measures her subsystem to obtain information about Bob's measurement result.}}
\centering
\end{figure*}

\section{Eavesdropper's strategies}
\indent For an intruder, there are two ways of eavesdropping. The first is to eavesdrop on the quantum state of sender Alice and the other is to eavesdrop on the result of receiver Bob. When the intruder  Eve, eavesdrops on the quantum state of sender Alice, she can do it using unambiguous discrimination, without an error. However, from the argument of sequential discrimination, this process can be observed by Alice and Bob \cite{m.namkung2}. Therefore, the sender and receiver can recognize the presence of an eavesdropper. \\
\indent When Eve wants to eavesdrop on the result of receiver Bob, she should be in a quantum entangled state with Bob. Assuming that the existence of an eavesdropper is unnoticed, the eavesdropping can be described as  a noisy quantum channel to of Alice and Bob as Fig. 1(a). 
When Alice prepares $|\psi_a\rangle$ ($a\in\{0,1\}$)
\begin{equation}
|\psi_a\rangle =\sqrt{\frac{1+s}{2}}|1\rangle+(-1)^a\sqrt{\frac{1-s}{2}}|2\rangle,
\end{equation}
with prior probability $q_a$, the noisy quantum channel between Alice and Bob can be described as follows:
\begin{equation}
\Lambda^{(A\rightarrow B)}(|\psi_a\rangle\langle\psi_a|)_A=\eta_{AB}|\psi_a\rangle\langle\psi_a|_B+(1-\eta_{AB})\frac{\mathbb{I}_B}{2}.\label{1}
\end{equation}
Here, the lower indices $A$ and $B$ denote the systems of Alice and Bob. $\mathbb{I}_B=|1\rangle\langle1|+|2\rangle\langle 2|$ is an identity operator defined in the system of Bob, which consists of an orthonormal basis $\{|1\rangle,|2\rangle\}$. In Eq. (\ref{1}), $\eta_{AB}\in[0,1]$ denotes the  channel efficiency between Alice and Bob.

\begin{figure*}[t]
\centerline{\includegraphics[width=14cm]{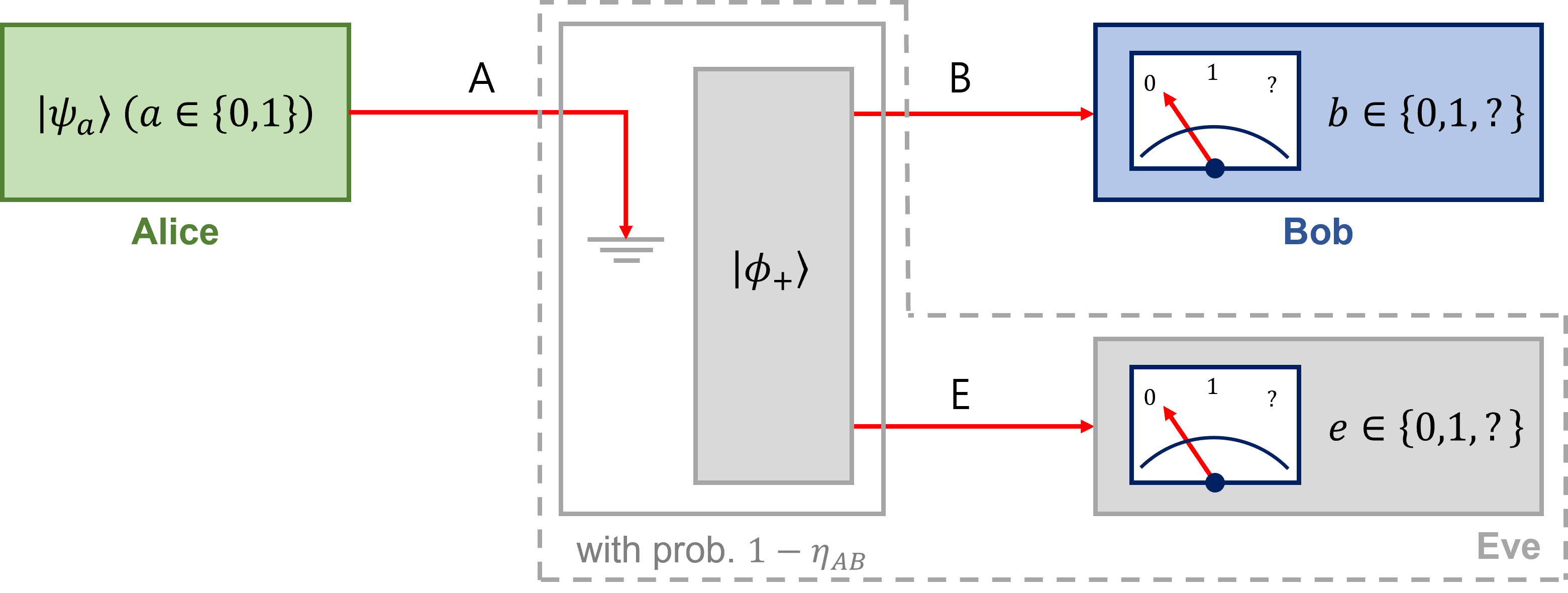}}
\caption{{Type-II structure of Eve's scheme. In this scheme, Eve lets Alice's state be transmitted to Bob with a probability $\eta_{AB}$, or discards Alice's state and shares $|\phi_+\rangle$ with Bob with a probability $1-\eta_{AB}$. Then, Eve performs a quantum measurement on her subsystem to discriminate Bob's measurement result $b$.}}
\centering
\end{figure*}

\subsection{Type-I structure of eavesdropper's scheme} Let us consider the eavesdropper's scheme illustrated as Fig. 1(b). If quantum systems of Bob and Eve are considered, Eve uses a quantum machine to deterministically transform the Alice's state $|\psi_a\rangle$ to a composite state between Bob and Eve:
\begin{equation}
|\Gamma_a\rangle_{BE}=\sqrt{\eta_{AB}}|\psi_a\rangle_B\otimes |0\rangle_E+\sqrt{1-\eta_{AB}}|\phi_+\rangle_{BE},\label{2}
\end{equation}
with an entangled state
\begin{equation}
|\phi_+\rangle_{BE}=\frac{1}{\sqrt{2}}(|11\rangle+|22\rangle)_{BE},\label{3}
\end{equation}
where is the entangled state between Bob and Eve. {Then, Eve performs a quantum measurement on her system to discriminate Bob's measurement result. If $\eta_{AB}$ is equal to one, then the composite state in Eq. (\ref{2}) is a product state. Thus, Eve cannot obtain information by measuring her subsystem. Otherwise, Eve can obtain the information about Bob's measurement result. We note that the partial state of Bob is equal to Eq. (\ref{1}).}

\subsection{Type-II structure of eavesdropper's scheme} 
The drawback of the eavesdropping scheme introduced above is that it requires a quantum machine deterministically producing $|\Gamma_a\rangle$. Since designing the quantum machine can be difficult, we further propose an alternative eavesdropping scheme. In this scheme, we can consider a composite state between Bob and Eve as follows:
\begin{equation}
\sigma_{a,BE}=\eta_{AB}|\psi_a\rangle\langle\psi_a|_B\otimes|0\rangle\langle0|_{E}+(1-\eta_{AB})|\phi_+\rangle\langle\phi_+|_{BE},\label{4}
\end{equation}
which satisfies $\mathrm{Tr}_E\sigma_{a,BE}=\Lambda^{(A\rightarrow B)}(|\psi_a\rangle\langle\psi_a|)$. {The procedure for producing the composite state in Eq. (\ref{4}) is illustrated in Fig. 2. In this figure, Eve lets Alice's state be transmitted to Bob with a probability $\eta_{AB}$, or discard Alice's state and share $|\phi_+\rangle$ with Bob with a probability $1-\eta_{AB}$.} 

\indent These two types can provide same security. That is because the joint measurement probability between Bob and Eve in the type-I structure is equal to that in the type-II structure (For detail, see Section 3). Particularly, the type-II structure can be easily reproduced in an experimental setup (For detail, see Section 4). 

\section{Sequential discrimination including eavesdropper}
{For the security analysis,} we propose the new sequential discrimination for describing the two eavesdropper's schemes. We first explain the structure of sequential discrimination, and propose the optimal success probability of eavesdropping. We further investigate the amount of the secret key rate in frame of the sequential discrimination scenario.

\subsection{Structure of sequential discrimination}
Let us first explain how each of the eavesdropping scheme introduced in the previous section is described as a sequential discrimination problem. It is noted that the unambiguous discrimination can be applied to the B92 protocol \cite{j.a.bergou,c.h.bennett}. For this reason, we consider that Bob has a quantum measurement which can unambiguously discriminates Alice's states $|\psi_0\rangle$ and $|\psi_1\rangle$.

\indent We first consider the type-I structure. We note in advance that our argument in here can also be applied to the type-II structure. Suppose that positive-operator valued measure (POVM) $\{M_0^{(B)},M_1^{(B)},M_?^{(B)}\}$ denotes the measurements of Bob. Then, the Kraus operator $K_b^{(B)}$ corresponding to the POVM element $M_b^{(B)}$ ($b\in\{0,1,?\}$) is given by   \cite{j.a.bergou3,m.namkung,m.namkung2}:
\begin{eqnarray}
&K_0^{(B)}=\sqrt{\alpha_0}|\phi_0^{(B)}\rangle\langle\alpha_0|, \ \ K_1^{(B)}=\sqrt{\alpha_1}|\phi_1^{(B)}\rangle\langle\alpha_1|,\nonumber\\
&K_?^{(B)}=\sqrt{1-\alpha_0}|\phi_0^{(B)}\rangle\langle\alpha_0|+\sqrt{1-\alpha_1}|\phi_1^{(B)}\rangle\langle\alpha_1|.\label{5}
\end{eqnarray}
Here, $\alpha_0$ and $\alpha_1$ are non-negative parameters  \cite{m.namkung2}, and $|\alpha_0\rangle$ and $|\alpha_1\rangle$ are corresponding vectors:
\begin{eqnarray}|\alpha_0\rangle&=&\frac{1}{\sqrt{2(1+s)}}|1\rangle+\frac{1}{\sqrt{2(1-s)}}|2\rangle,\nonumber\\
|\alpha_1\rangle&=&\frac{1}{\sqrt{2(1+s)}}|1\rangle-\frac{1}{\sqrt{2(1-s)}}|2\rangle.\label{a0a1new}
\end{eqnarray}
For $a\not=b$, the inner product between $|\alpha_b\rangle$ and $|\psi_a\rangle$ is equal to zero. It guides us to the fact that the measurement described in terms of the Kraus operators in Eq. (\ref{5}) can perform the unambiguous discrimination. When Bob obtains a  conclusive result $b\in\{0,1\}$, the Kraus operator $K_b^{(B)}$ probabilistically changes the bipartite state of Eq. (\ref{2}) into the following form:
\begin{eqnarray}
K_0^{(B)}\otimes\mathbb{I}_E|\Gamma_0\rangle_{BE}&=&|\phi_0^{(B)}\rangle_B\otimes|\gamma_{00}\rangle,\nonumber\\
K_1^{(B)}\otimes\mathbb{I}_E|\Gamma_0\rangle_{BE}&=&|\phi_1^{(B)}\rangle_B\otimes|\gamma_{01}\rangle,\nonumber\\
K_0^{(B)}\otimes\mathbb{I}_E|\Gamma_1\rangle_{BE}&=&|\phi_0^{(B)}\rangle_B\otimes|\gamma_{10}\rangle,\nonumber\\
K_1^{(B)}\otimes\mathbb{I}_E|\Gamma_1\rangle_{BE}&=&|\phi_1^{(B)}\rangle_B\otimes|\gamma_{11}\rangle,\nonumber\\ \label{kr}
\end{eqnarray}
where $|\gamma_{ab}\rangle$ are written as
\begin{eqnarray}
|\gamma_{00}\rangle&=&\mathcal{N}\left\{\sqrt{\eta_{AB}\alpha_0}|0\rangle_E+\sqrt{\frac{(1-\eta_{AB})\alpha_0}{2(1-s^2)}}|\widetilde{\psi}_0\rangle_E\right\},\nonumber\\
|\gamma_{01}\rangle&=&|\widetilde{\psi}_1\rangle_E,\nonumber\\
|\gamma_{10}\rangle&=&|\widetilde{\psi}_0\rangle_E,\nonumber\\
|\gamma_{11}\rangle&=&\mathcal{N}\left\{\sqrt{\eta_{AB}\alpha_1}|0\rangle_E+\sqrt{\frac{(1-\eta_{AB})\alpha_1}{2(1-s^2)}}|\widetilde{\psi}_1\rangle_E\right\}.
\end{eqnarray}
Here, $\mathcal{N}$ is the normalization constant and 
\begin{equation}
|\widetilde{\psi}_b\rangle=\sqrt{1-s^2}|\alpha_b\rangle\label{add1}
\end{equation} 
\noindent is a pure state spanned by $\{|1\rangle,|2\rangle\}$. According to Eq. (\ref{add1}), $|\widetilde{\psi}_b\rangle$ is orthogonal to $|0\rangle$. Moreover, the label of $|\widetilde{\psi}_b\rangle$ in Eq. (\ref{kr}) is equal to the measurement result of Bob. Therefore, Eve can eavesdrop the measurement result of Bob by discriminating $|\widetilde{\psi}_0\rangle$ and $|\widetilde{\psi}_1\rangle$ with her measurement described as the POVM $\{M_0^{(E)},M_1^{(E)},M_?^{(E)}\}$ on the subspace spanned by $\{|1\rangle,|2\rangle\}$,
\begin{eqnarray}
M_0^{(E)}&=&u_0|u_0\rangle\langle u_0|, \nonumber \\ 
M_1^{(E)}&=&u_1|u_1\rangle\langle u_1|, \nonumber \\ 
M_?^{(E)}&=&\mathbb{I}_E-M_0^{(E)}-M_1^{(E)},\label{povm_eee}
\end{eqnarray}
\noindent where $M_e^{(E)}$ is the POVM element corresponding to the measurement result $e$. In Eq. (\ref{povm_eee}), $\mathbb{I}_E$ is the identity operator on Eve's system, $u_e$ is the non-negative real number, and $|u_e\rangle$ is the vector in the subspace $\{|1\rangle,|2\rangle\}$ satisfying $\langle\widetilde{\psi}_b|u_e\rangle=\delta_{be}$. We note that $|u_e\rangle$ can be constructed in the same way as Eq. (\ref{a0a1new}) \cite{m.namkung2}. 

\begin{figure}[t]
\centerline{\includegraphics[width=15cm]{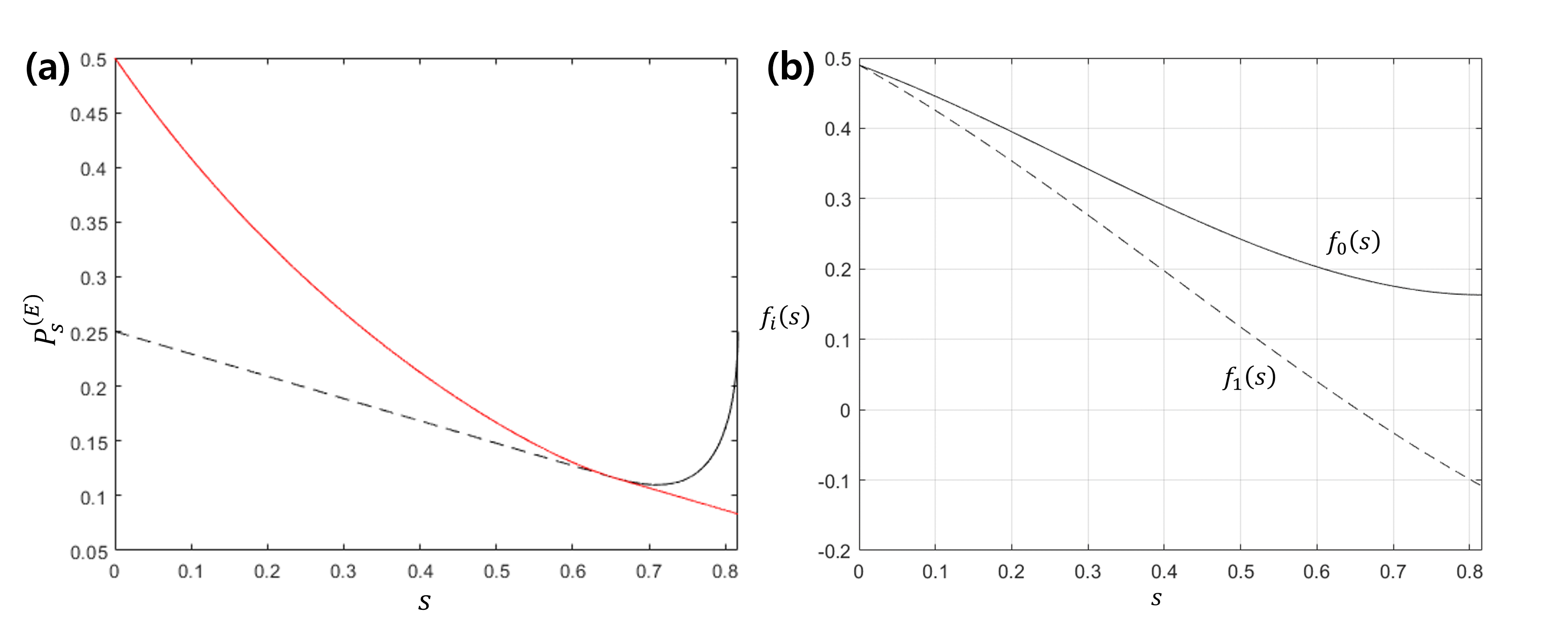}}
\caption{(a) Success probability of eavesdropping. Solid black line and dashed black line are $P_{s,opt1}^{(E)}$ and $P_{s,opt2}^{(E)}$ in Eq. (\ref{19_}), respectively, and solid red line is the optimal success probability of eavesdropping. In (b), $f_0(s)$ and $f_1(s)$ in Eq. (\ref{fff}) are depicted.}
\centering
\end{figure}

\indent In the aspect of the quantum state discrimination task, the finite (but nonzero) success probability implies that a receiver can obtain an information about sender's state \cite{j.a.bergou}. Thus, one of the probable figures of merit is ``the success probability of eavesdropping'' in case of type-I structure, which is described as (the detailed evaluation is presented in Appendix A.1)
\begin{eqnarray}
P_{s,\mathrm{type-I}}^{(E)}=\sum_{a,b\in\{0,1\}}q_a\langle\Gamma_a|K_b^{(B)\dagger}K_b^{(B)}\otimes\mathbb{I}_E|\Gamma_a\rangle\langle\gamma_{ab}|M_b^{(E)}|\gamma_{ab}\rangle. \label{11}
\end{eqnarray}
Assume that Bob performs optimal unambiguous discrimination on Alice's state. Then, $P_{s,opt}^{(E)}$, which is the optimum success probability of eavesdropping, can have a simple expression such as $P_{s,opt1}^{(E)}$ or $P_{s,opt2}^{(E)}$, 
\begin{eqnarray}
P_{s,opt}^{(E)}&=&\frac{1-\eta_{AB}}{2(1-s^2)}(\alpha_0+\alpha_1-2\sqrt{\alpha_0\alpha_1}s),  \ \  \mathrm{if} \ \  f_0(s)>0 \ \  \mathrm{and} \ \ f_1(s)>0,\nonumber
      \\
      P_{s,opt}^{(E)}&=&\frac{1-\eta_{AB}}{2}\max\{\alpha_0,\alpha_1\},
        \ \   \mathrm{if} \ \  f_0(s)\le0 \ \ \mathrm{or} \ \  f_1(s)\le0,
\label{19_}
\end{eqnarray}
with $s:=|\langle\psi_1|\psi_2\rangle|$ and
\begin{eqnarray}
&&f_0(s):=q_1s^3-\sqrt{q_0q_1}s^2-q_0s+\sqrt{q_0q_1},\nonumber\\
&&f_1(s):=q_0s^3-\sqrt{q_0q_1}s^2-q_1s+\sqrt{q_0q_1}.\label{fff}
\end{eqnarray}
\noindent The detailed evaluation of the optimization is presented in the Appendix A.2. If $s\in[0,\sqrt{q_1/q_2}]$, we get $\alpha_0=1-\sqrt{\frac{q_1}{q_0}}s$ and $\alpha_1=1-\sqrt{\frac{q_0}{q_1}}s$ from Bob's optimal POVM condition \cite{g.jaeger}. 

\indent Fig. 3(a) illustrates the optimum success probability of eavesdropping($P_{s,opt}^{(E)}$) in Eq. (\ref{19_}). Here, we have used $q_0=0.4(q_1=0.6)$ and $\eta_{AB}=0.5$. In Fig. 3(a), the solid black line(dashed black line) indicates $P_{s,opt1}^{(E)}$ ($P_{s,opt2}^{(E)}$). According to Fig. 3(a), in the region of $s<0.6538$, $P_{s,opt1}^{(E)}$ (solid black line) is optimum. That is because, as illustrated in Fig. 3(b), both $f_0(s)$ and $f_1(s)$ in Eq. (\ref{fff}) are non-negative in this region. Meanwhile, $P_{s,opt2}^{(E)}$ (dashed black line) is optimum in the region of $s>0.6538$, since one of $f_1(s)$ is negative. Thus, the optimum success probability of eavesdropping is indicated by the solid red line.

We further evaluate the success probability of eavesdropping in type-II structure as
\begin{eqnarray}
P_{s,\mathrm{type-II}}^{(E)}=\sum_{a,b\in\{0,1\}}q_a\mathrm{tr}\left[K_b^{(B)}\otimes\mathbb{I}_E\sigma_{a,BE}K_b^{(B)\dagger}\otimes\mathbb{I}_E\right]\mathrm{tr}\left[\tau_{ab,E}M_b^{(E)}\right],\label{soe_mix_}
\end{eqnarray}
where $\tau_{ab,E}$ are defined as
\begin{eqnarray}
&&\tau_{00,E}=\frac{\eta_{AB}}{\eta_{AB}+\frac{1-\eta_{AB}}{2(1-s^2)}}|0\rangle\langle0|_E+\frac{\frac{1-\eta_{AB}}{2(1-s^2)}}{\eta_{AB}+\frac{1-\eta_{AB}}{2(1-s^2)}}|\widetilde{\psi}_0\rangle\langle\widetilde{\psi}_0|_E,\nonumber\\
&&\tau_{01,E}=|\widetilde{\psi}_1\rangle\langle\widetilde{\psi}_1|_E,\nonumber\\
&&\tau_{10,E}=|\widetilde{\psi}_0\rangle\langle\widetilde{\psi}_0|_E,\nonumber\\
&&\tau_{11,E}=\frac{\eta_{AB}}{\eta_{AB}+\frac{1-\eta_{AB}}{2(1-s^2)}}|0\rangle\langle0|_E+\frac{\frac{1-\eta_{AB}}{2(1-s^2)}}{\eta_{AB}+\frac{1-\eta_{AB}}{2(1-s^2)}}|\widetilde{\psi}_1\rangle\langle\widetilde{\psi}_1|_E. \label{tau_}
\end{eqnarray}
From the straightforward calculation, the success probability of eavesdropping in Eq. (\ref{soe_mix_}) is equal to Eq. (\ref{11}). The proof is presented in Appendix A.3. Thus, the optimal success probability of eavesdropping in type-II structure is also analytically derived as Eq. (\ref{19_}).

\begin{figure*}[t]
\centerline{\includegraphics[width=15cm]{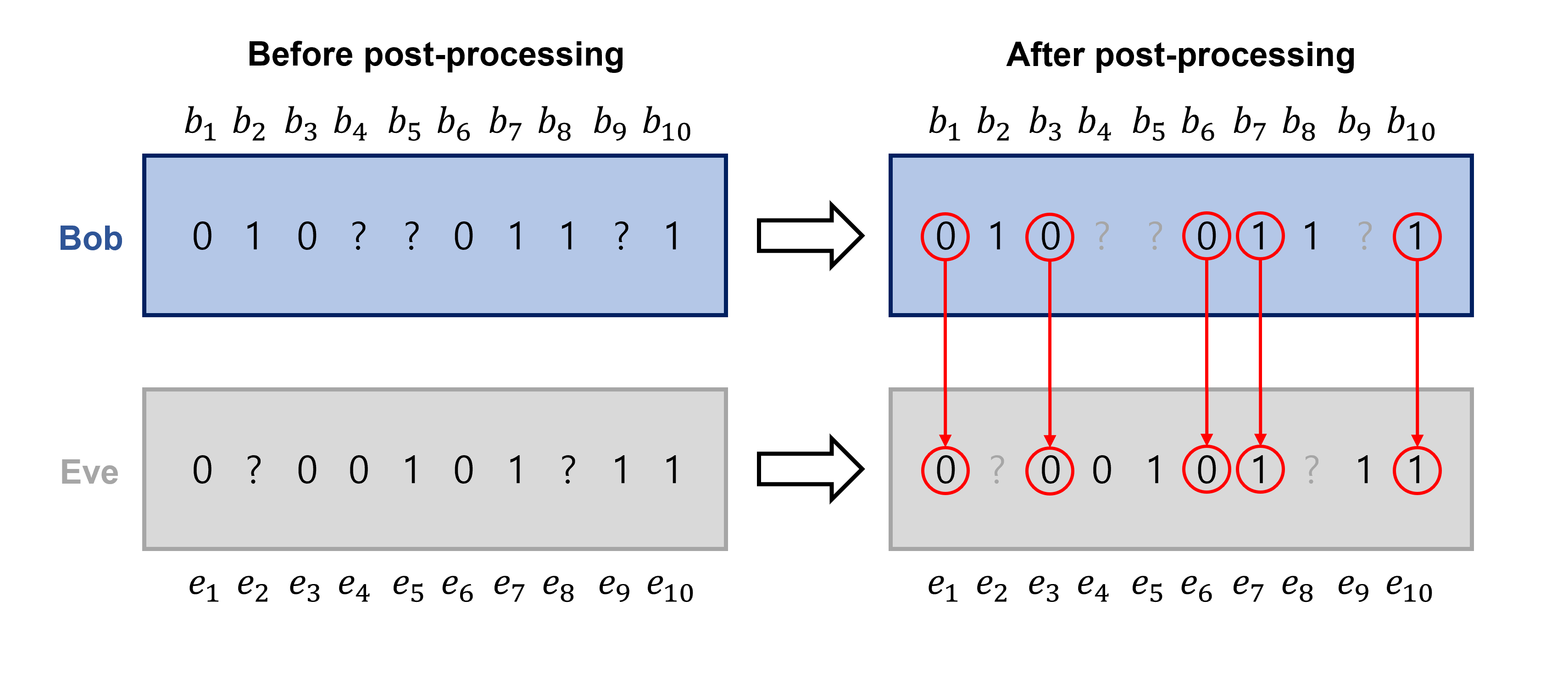}}
\caption{Post-processing performed by Bob and Eve. Let us suppose that Bob has 10 measurement results $b_1,\cdots,b_{10}$, and Eve has measurement results $e_1,\cdots,e_{10}$. Bob can discard the inconclusive results $b_4,b_5,b_{10}$, and Eve can also discard $e_2$ and $e_8$.}
\centering
\end{figure*}
\begin{figure}[t]
\centerline{\includegraphics[width=9.5cm]{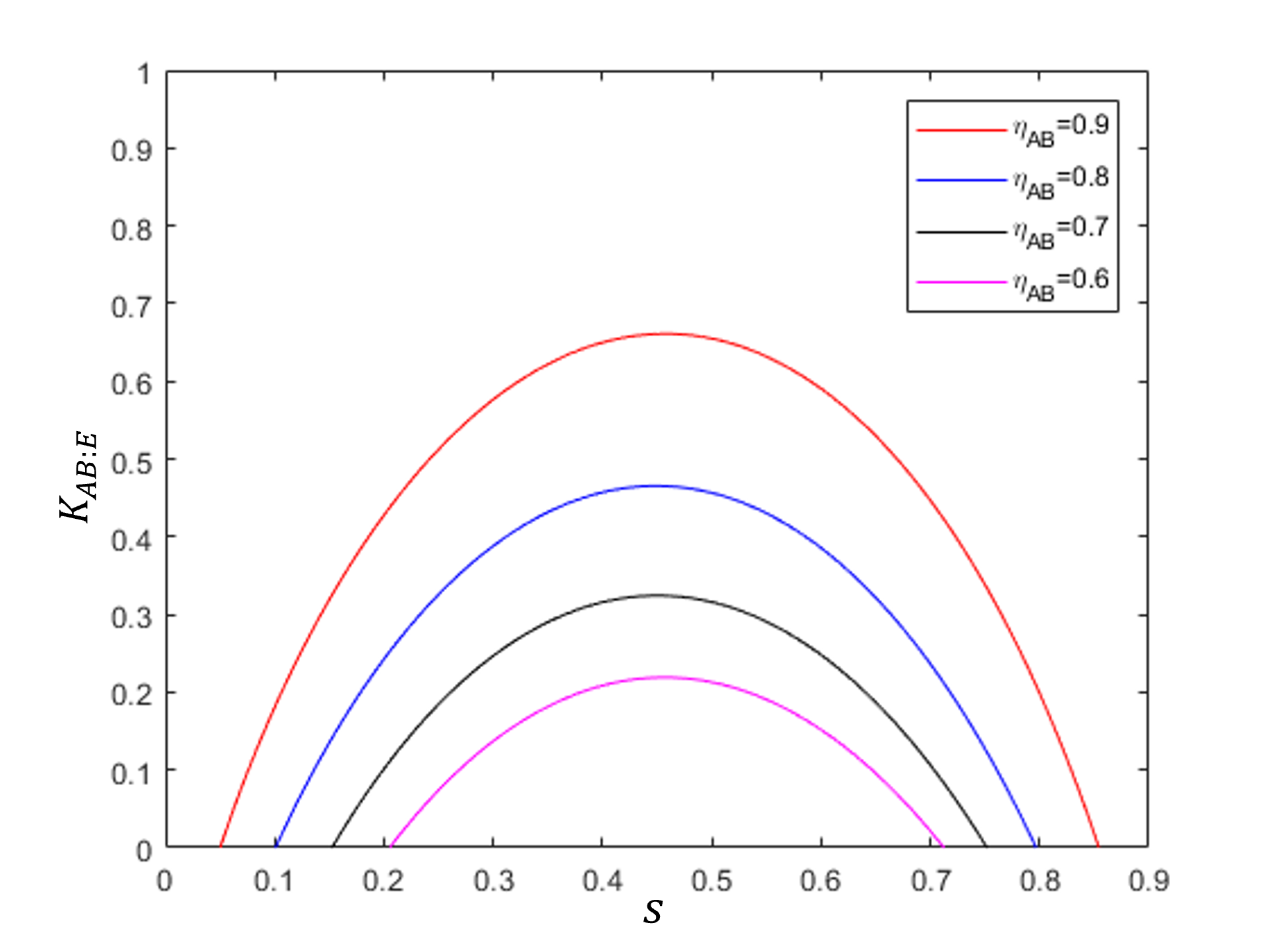}}
\caption{Secret key rate $K_{AB:E}$: red, blue, black, and purple lines correspond to $\eta_{AB}=0.9$, $\eta_{AB}=0.8$, $\eta_{AB}=0.7$, and $\eta_{AB}=0.6$, respectively.}
\centering
\end{figure}

\subsection{Secret key rate}
According to Csiszar and Korner \cite{i.csiszar}, when the amount of information between a receiver and sender is larger than that between a receiver and eavesdropper, a secret key can exist as an amount equal to the difference of information. The secret key rate is defined as
\begin{eqnarray}
K_{AB:E}&=&\max\{0,I(B:A)-I(B:E)\}\nonumber\\
&=&\max\{0,H(A)-H(B,A)-H(E)+H(B,E)\}. \label{21}
\end{eqnarray}
Here, $I(X:Y)=H(X)+H(Y)-H(X,Y)$ is Shannon mutual information. $H(X)$ denotes Shannon entropy and $H(X,Y)$ is Shannon joint entropy. If $K_{AB:E}>0$, sender Alice and receiver Bob can share the secret key \cite{i.csiszar}.\\
\indent As illustrated in Fig. 4, Bob and Eve can perform the following post-processing. In case that Bob performs optimal unambiguous discrimination, he can discard the measurement result when he obtains an inconclusive result. This post-processing can enhance the amount of information shared between Alice and Bob \cite{d.fields}. In this way, the joint probability between Alice and Bob is 
\begin{equation}
\widetilde{P}_{AB}(a,b)=\frac{P_{AB}(a,b)}{\sum_{a,b\in\{0,1\}}P_{AB}(a,b)},
\end{equation}
which constitutes the Shannon mutual information in Eq. (\ref{21}). Here. $a,b\in\{0,1,?\}$ are the measurement results for Alice and Bob, respectively. Similarly, when Eve obtains an inconclusive result, she discards the measurement result. Thus, it seems that Eve can successfully obtain information about Bob. However, Bob and Eve are separated in space and the information leakage discussed above is not permitted. In other words, Eve cannot discard  her measurement result based on whether Bob obtained an inconclusive result or not. Therefore, the joint probability between Bob and Eve should be changed as follows: 
\begin{eqnarray}
\widetilde{P}_{BE}(b,e)=\frac{P_{BE}(b,e)}{\sum_{b\in\{0,1,?\}}\sum_{e\in\{0,1\}}P_{BE}(b,e)},\label{post}
\end{eqnarray}
where $b,e\in\{0,1,?\}$ are the measurement results for Bob and Eve, respectively.

Fig. 5 shows the secret key rate $K_{AB:E}$, considering the marginal probability between Bob and Eve which is updated from Eq. (\ref{post}). We note that the both two types of eavesdropper's scheme provides same secret key rate (for detail, see Appendix B). Here, the channel efficiency is considered as $\eta_{AB}=0.9$(solid red line), $\eta_{AB}=0.8$(solid blue line), $\eta_{AB}=0.7$(solid black line), and $\eta_{AB}=0.6$(solid purple line). 
As shown in Fig. 5, as the overlap $s$ increases, $K_{AB:E}$ also increases. However, from a specific overlap $K_{AB:E}$ decreases. For example, for $\eta_{AB}=0.9$, in the region of $s<0.4585$, $K_{AB:E}$ increases but in the region of $s>0.4585$, $K_{AB:E}$ decreases.

 The secret key rate $K_{AB:E}$ exhibits interesting behavior. When the overlap $s$ is large, it is difficult for Bob and Eve to efficiently implement QSD. In this case, the mutual information between Alice and Bob, and Bob and Eve becomes small. However, when $s$ is small, Bob and Eve can easily and efficiently implement QSD. In this case, the mutual information between Alice and Bob, and Bob and Eve becomes large.  

\begin{figure*}[t]
\centerline{\includegraphics[width=13cm]{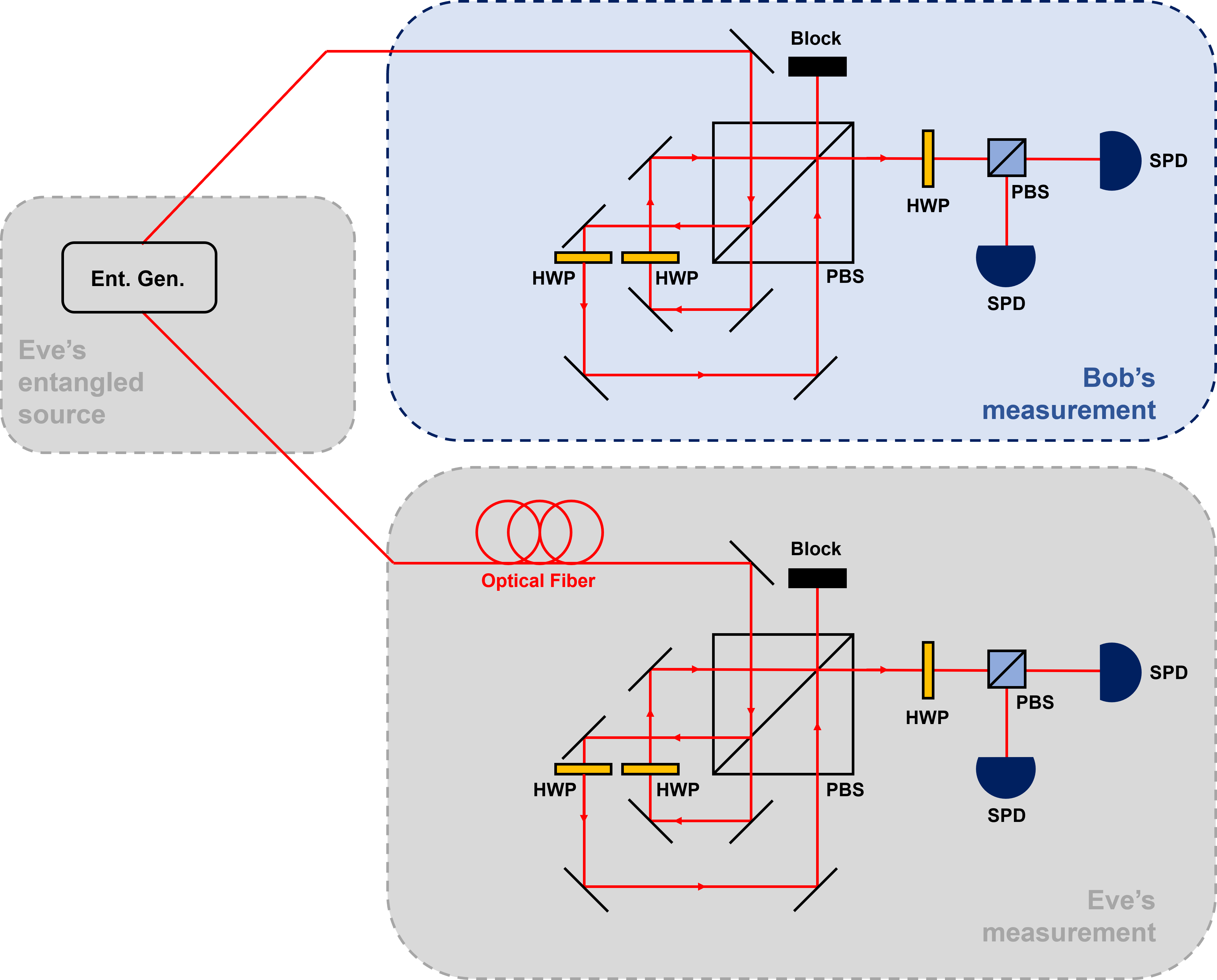}}
\caption{Experimental setting for Eve's eavesdropping. Here, Eve prepares maximally entangled state $|\phi_+\rangle=\frac{1}{\sqrt{2}}(|hh\rangle+|vv\rangle)$ between Bob and Eve with  probability $1-\eta_{AB}$. HWP: half-wave plate, PBS: polarized beam splitter, SPD: single-photon detector, and Ent. Gen.: entanglement generator \cite{p.g.kwait}.}
\centering
\end{figure*}

\section{Method for experimental implementation}
Let us propose an experimental method for a unified model of sequential state discrimination including an eavesdropper with quantum optics. Even though the type-I structure was used previously, we will use type-II structure, because it can be easily implemented in an experimental setup. In the type-II structure, Alice prepares a quantum state
\begin{equation}
|\psi_a\rangle=\sqrt{\frac{1+s}{2}}|h\rangle+(-1)^a\sqrt{\frac{1-s}{2}}|v\rangle,
\end{equation}
where $|h\rangle$ and $|v\rangle$ represent horizontal and vertical directions, respectively. Eve, who controls channel efficiency $\eta_{AB}$, can eavesdrop as follows: (i) With a probability of $\eta_{AB}$, Eve does not eavesdrop on the quantum state of Alice. (ii) With a probability of $1-\eta_{AB}$, Eve eliminates the quantum state of Alice and shares a maximally entangled state with Bob. (iii) After Bob’s measurement, Eve performs measurement on her subsystem.\\
\indent In Fig. 6 of the next page, we illustrate the experimental setup(for details about the description, see Appendix C). Here, the experimental setup of Bob and Eve is based on a Sagnac-like interferometer  \cite{f.a.torres-ruiz}. The setup consists of a half-wave plate(HWP), polarized beam splitter(PBS), and single-photon detector(SPD). In step (ii), Eve generates a maximally entangled two-polarization state $|\phi_+\rangle=\frac{1}{\sqrt{2}}(|hh\rangle+|vv\rangle)$, using a type-II spontaneous parametric down conversion(SPDC) \cite{p.g.kwait}. Type-II SPDC includes beta-barium borate(BBO) crystals, two birefringent  crystals, HWP, and quarter-wave plate(QWP). HWP and QWP transform the entangled pure state, generated by the BBO and birefringent crystals, into one of the four Bell-states. 

According to the type-II structure, if Eve generates $|\phi_+\rangle$ with a probability of $1-\eta_{AB}$, Eve can eavesdrop on the result of Bob, based on the selection of the path of a single photon and the measurement result of two SPDs. Ideally, Bob performs an unambiguous discrimination based on a Sagnac-like interferometer, and Eve can eavesdrop with the optimum success probability of eavesdropping by constructing a Sagnac-like interferometer. It should be emphasized that despite the attack by Eve, Alice and Bob can obtain the secret key rate.\\
\indent In reality, one should consider imperfections occurring in the photon state and in SPD. We consider the dark count rate($\nu>0$) and detection efficiency($0<\eta<1$) for the SPD. The photon state in the setup consists of two types: a single-photon polarization state that Alice sends to Bob, and the single photon state of maximally entangled state generated by Eve. Different types of  photon states suffer from different types of noises. For example, the single-photon polarization state may disappear under a noisy channel, which is called ``amplitude damping''  \cite{m.a.nielson,y.-s.kim}. We assume that amplitude damping can occur between Alice and Bob and between Bob and Eve. In addition, white or colored noise can occur when Eve generates a maximally entangled quantum state \cite{a.cabello}. Particularly, colored noise which occurs because of imperfections in experimental entangling operations is more frequent than white noise \cite{a.cabello}.\\

\begin{figure*}[t]
\centerline{\includegraphics[width=14cm]{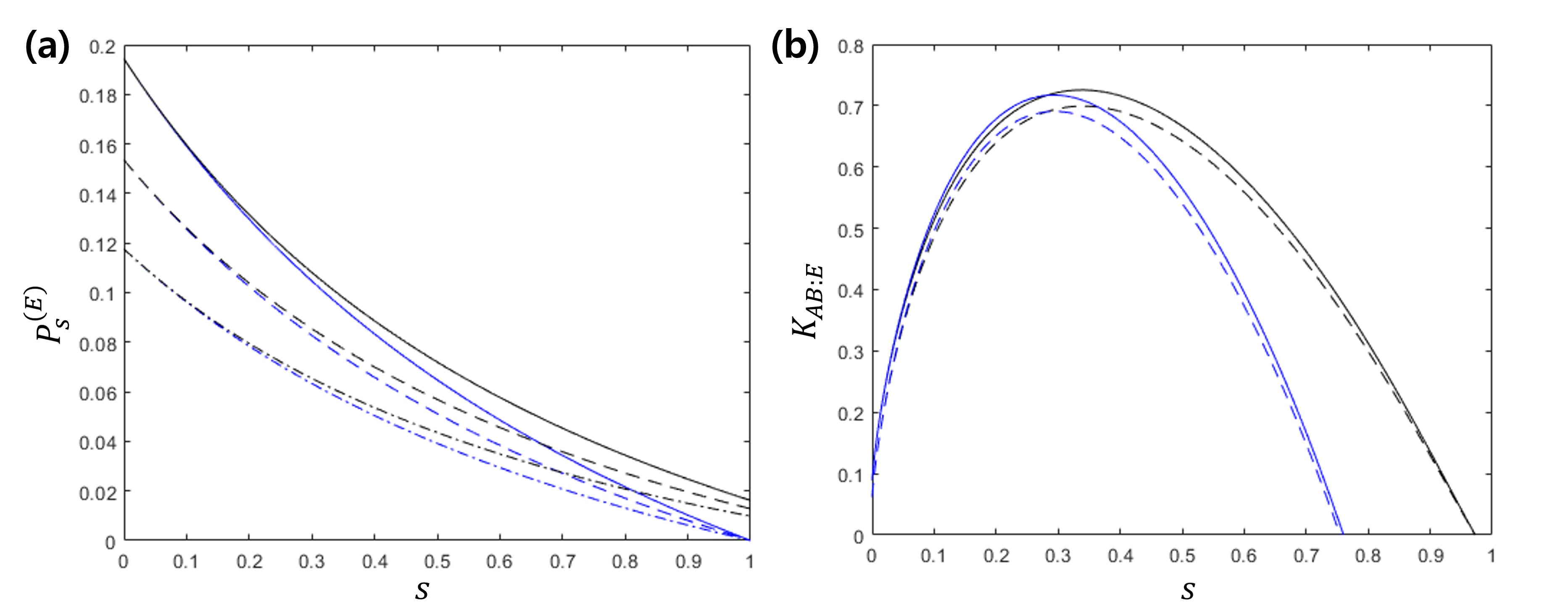}}
\caption{(a) Success probability of eavesdropping under imperfect quantum channel, entangled state, and single-photon detector. (b) Secret key rate between Alice and Bob. Here, $\eta_{AB}=0.5$, $\eta_{ent}=0.5$, and $\eta=0.8$ are considered. Blue(black) line corresponds to color(white) noise. Solid, dashed, and dash-dot lines correspond to $D=0.1$, $D=0.2$, and $D=0.3$, respectively. {Here, we assume that $D_0=D_e=D$ for considering the relation between the secret key rate and a single decoherence parameter. Secret key rate under imperfect quantum channel, entangled state, and single-photon detector. Here, $\eta_{AB}=0.5$, $\eta_{ent}=0.5$, and $\eta=0.8$ are considered. Blue(black) line corresponds to color(white) noise. Solid(dashed) line corresponds to $D_0=0.1$($D_0=0.2$). In every case, $D_e=0.4$ is considered.}}
\centering
\end{figure*}

\indent The success probability of eavesdropping under white and colored noise is displayed in Fig. 7(a) (for detail, see Appendix D). In Fig. 7(a), the value of $\eta_{ent}=0.5$, $\eta=0.8$,  and $\eta_{AB}=0.5$ are considered, where the detection efficiency $\eta=0.8$ is the value of a commercialized superconducting nanowire single-photon detector(SNSPD) whose dark count rate is nearly zero \cite{snspd}. In Fig. 7(a), the solid line, dashed  line, and dash-dot line correspond to the cases of decoherence parameter, $D=0.1$, $D=0.2$, and $D=0.3$, respectively(a large $D$ implies that the decoherence rate is high). {Here, we assume that $D_0=D_e=D$ for considering the relation between the secret key rate and a single decoherence parameter.} The black and blue lines show the cases of white and colored noise, respectively. 

\indent In Fig. 7(b), the secret key rate between Alice and Bob is displayed, considering various imperfections (for detail, see Appendix D). Here, $\eta_{AB}=0.5$, $\eta_{ent}=0.5$, and $\eta=0.8$ are considered. The blue(black) line corresponds to colored(white) noise. The solid(dashed) line corresponds to $D_0=0.1$($D_0=0.2$). In every case, $D_e$ is taken as $0.4$. It should be  noted that the secret key rate does not change when $D_0=D_e$ owing to the post-processing expressed in Eq. (\ref{post}). As shown in Fig. 7(b), the graph of the secret key rate has one global maximum. This implies that (i) if $s$ tends to be smaller, then the secret key rate decreases because the tendency of $s$ makes Eve as well as Bob to easily discriminate the quantum states, and (ii) if $s$ tends to be larger, then the secret key rate decreases because the tendency of $s$ makes discrimination performed by Bob and Eve difficult. 
\\
\indent For both Fig. 7(a) and (b), we observe that the success probability of eavesdropping can be smaller in the case of the colored noise than in the case of white noise. These results {contradict} our previously held belief. This is because the colored noise preserves the probabilistic correlation between Bob and Eve, whereas white noise does not.

\section{Conclusion}
In this paper, we have proposed a unified model of sequential state discrimination including an eavesdropper. We have shown that even though Eve uses an entanglement to eavesdrop on Bob's measurement result, Alice and Bob can have a non-zero secret key rate. Furthermore, we have proposed an experimental model for eavesdropping. Because our experimental method consists of linear optical technologies, the implementation of our method is practical. Ideally, our experiment can achieve optimum success probability of eavesdropping. And we have investigated possible imperfections including quantum channels between Alice and Bob, entanglement between Bob and Eve, and the inefficiency of Bob's SPD. Remarkably, under these imperfections, we have shown that the success probability of eavesdropping in the case of colored  noise can be smaller than those in the case of white noise.

{It should be noted that our sequential discrimination model can be extended to the case of unambiguously discriminating $N$ pure states \cite{a.chefles,j.a.bergou2}. This extension is important since large $N$ guarantees large amount of transmitted bits per a signal pulse. Moreover, our experimental idea can also be applied to the continuous variable version. That is because sequential measurement that unambiguously discriminates two coherent states can be designed with linear optics \cite{m.namkung3}.} \\ 

\section*{Acknowledgements}
This work is supported by the Basic Science Research Program through the National Research Foundation of Korea (NRF) funded by the Ministry of Education, Science and Technology (NRF2020M3E4A1080088 and NRF2022R1F1A1064459) and and Creation of the Quantum Information Science RD Ecosystem (Grant No. 2022M3H3A106307411) through the National Research Foundation of Korea (NRF) funded by the Korean government (Ministry of Science and ICT).

\bibliographystyle{plain}

\onecolumn\newpage
\appendix

\section*{Appendix A. Success probability of eavesdropping}
In this section, we derive and optimize the success probability of eavesdropping by considering both types of eavesdropping strategies.

\subsection*{A.1. Describing type-I structure of eavesdropper}
In this structure, the following entangled state between Bob and Eve is considered:
\begin{equation}
|\Gamma_a\rangle_{BE}=\sqrt{\eta_{AB}}|\psi_a\rangle_B\otimes|0\rangle_E+\sqrt{\frac{1-\eta_{AB}}{2}}(|11\rangle+|22\rangle)_{BE}.\label{gamma_i}
\end{equation}
Then, each $K_b^{(B)}\otimes\mathbb{I}_E|\Gamma_a\rangle$ are obtained by
\begin{eqnarray}
K_0^{(B)}\otimes\mathbb{I}_E|\Gamma_0\rangle&=&\sqrt{\eta_{AB}\alpha_0}|\phi_0^{(B)}\rangle_B\otimes|0\rangle_E\nonumber\\&+&\sqrt{\frac{(1-\eta_{AB})\alpha_0}{2}} \left\{(|\phi_0^{(B)}\rangle\langle \alpha_0|\otimes\mathbb{I}_E)|11\rangle+(|\phi_0^{(B)}\rangle\langle \alpha_0|\otimes\mathbb{I}_E)|22\rangle \right\}_{BE},\nonumber\\
K_1^{(B)}\otimes\mathbb{I}_E|\Gamma_0\rangle&=&\sqrt{\frac{(1-\eta_{AB})\alpha_1}{2}}\left\{(|\phi_1^{(B)}\rangle\langle \alpha_1|\otimes\mathbb{I}_E)|11\rangle+(|\phi_1^{(B)}\rangle\langle \alpha_1|\otimes\mathbb{I}_E)|22\rangle\right\}_{BE},\nonumber\\
K_0^{(B)}\otimes\mathbb{I}_E|\Gamma_1\rangle&=&\sqrt{\frac{(1-\eta_{AB})\alpha_0}{2}}
\left\{(|\phi_0^{(B)}\rangle\langle \alpha_0|\otimes\mathbb{I}_E)|11\rangle+(|\phi_0^{(B)}\rangle\langle \alpha_0|\otimes\mathbb{I}_E)|22\rangle\right\}_{BE},\nonumber\\
K_1^{(B)}\otimes\mathbb{I}_E|\Gamma_1\rangle&=&\sqrt{\eta_{AB}\alpha_1}|\phi_1^{(B)}\rangle_B\otimes|0\rangle_E\nonumber\\
&+&\sqrt{\frac{(1-\eta_{AB})\alpha_1}{2}}\left\{(|\phi_1^{(B)}\rangle\langle \alpha_1|\otimes\mathbb{I}_E)|11\rangle+(|\phi_1^{(B)}\rangle\langle \alpha_1|\otimes\mathbb{I}_E)|22\rangle\right\}_{BE}.\nonumber\\ \label{k_gamma}
\end{eqnarray}
Without loss of generality, we write pure states $|\psi_a\rangle$ as
\begin{eqnarray}
|\psi_0\rangle_E&=&\sqrt{\frac{1+s}{2}}|1\rangle_E+\sqrt{\frac{1-s}{2}}|2\rangle_E,\nonumber\\
|\psi_1\rangle_E&=&\sqrt{\frac{1+s}{2}}|1\rangle_E-\sqrt{\frac{1-s}{2}}|2\rangle_E,\label{psi}
\end{eqnarray}
and vectors $|\alpha_b\rangle$ in the Kraus operators as
\begin{eqnarray}
|\alpha_0\rangle_E&=&\frac{1}{\sqrt{2(1+s)}}|1\rangle_E+\frac{1}{\sqrt{2(1-s)}}|2\rangle_E,\nonumber\\
|\alpha_1\rangle_E&=&\frac{1}{\sqrt{2(1+s)}}|1\rangle_E-\frac{1}{\sqrt{2(1-s)}}|2\rangle_E,\label{alpha}
\end{eqnarray}
such that $\langle\psi_a|\alpha_b\rangle=\delta_{ab}$ for every $a,b\in\{0,1\}$. Substituting Eq. (\ref{psi}) and Eq. (\ref{alpha}) into Eq. (\ref{k_gamma}), we obtain
\begin{eqnarray}
K_0^{(B)}\otimes\mathbb{I}_E|\Gamma_0\rangle&=&\sqrt{\eta_{AB}\alpha_0}|\phi_0^{(B)}\rangle_B\otimes|0\rangle_E+\sqrt{\frac{(1-\eta_{AB})\alpha_0}{2}}|\phi_0^{(B)}\rangle_B\otimes|\alpha_0\rangle_E,\nonumber\\
K_1^{(B)}\otimes\mathbb{I}_E|\Gamma_0\rangle&=&\sqrt{\frac{(1-\eta_{AB})\alpha_1}{2}}|\phi_1^{(B)}\rangle_B\otimes|\alpha_1\rangle_E,\nonumber\\
K_0^{(B)}\otimes\mathbb{I}_E|\Gamma_1\rangle&=&\sqrt{\frac{(1-\eta_{AB})\alpha_0}{2}}|\phi_0^{(B)}\rangle_B\otimes|\alpha_0\rangle_E,\nonumber\\
K_1^{(B)}\otimes\mathbb{I}_E|\Gamma_1\rangle&=&\sqrt{\eta_{AB}\alpha_1}|\phi_1^{(B)}\rangle_B\otimes|0\rangle_E+\sqrt{\frac{(1-\eta_{AB})\alpha_1}{2}}|\phi_1^{(B)}\rangle_B\otimes|\alpha_1\rangle_E.
\label{rewritten_k_gamma}
\end{eqnarray}
We define (normalized) pure states by
\begin{equation}
|\widetilde{\psi}_a\rangle_E=\sqrt{1-s^2}|\alpha_a\rangle_E.\label{tilde_psi}
\end{equation}
Substituting Eq. (\ref{tilde_psi}) into Eq. (\ref{rewritten_k_gamma}), we obtain the representation of Eq. (6) in the letter with $|\gamma_{ab}\rangle$ defined by
\begin{eqnarray}
|\gamma_{00}\rangle_E&=&\frac{1}{\sqrt{\eta_{AB}+\frac{(1-\eta_{AB})}{2(1-s^2)}}}\left\{\sqrt{\eta_{AB}}|0\rangle+\sqrt{\frac{(1-\eta_{AB})}{2(1-s^2)}}|\widetilde{\psi}_0\rangle\right\}_E,\nonumber\\
|\gamma_{01}\rangle_E&=&|\widetilde{\psi}_1\rangle_E,\nonumber\\
|\gamma_{10}\rangle_E&=&|\widetilde{\psi}_0\rangle_E,\nonumber\\
|\gamma_{11}\rangle_E&=&\frac{1}{\sqrt{\eta_{AB}+\frac{(1-\eta_{AB})}{2(1-s^2)}}}\left\{\sqrt{\eta_{AB}}|0\rangle+\sqrt{\frac{(1-\eta_{AB})}{2(1-s^2)}}|\widetilde{\psi}_1\rangle\right\}_E.\label{gamma_vector}
\end{eqnarray}

Consider Eve's POVM as $\{M_0^{(E)},M_1^{(E)},M_?^{(E)}\}$ where each POVM element is given by
\begin{equation}
M_0^{(E)}=u_0|u_0\rangle\langle u_0|, \ \ M_1^{(E)}=u_1|u_1\rangle\langle u_1|, \ \ M_?^{(E)}=\mathbb{I}_E-M_0^{(E)}-M_1^{(E)},\label{povm_e}
\end{equation}
where $u_0$ and $u_1$ are non-negative real numbers, and $|u_0\rangle$ and $|u_1\rangle$ are vectors orthogonal to $|0\rangle_E$ and satisfying $\langle u_b|\widetilde{\psi}_a\rangle=\delta_{ab}$ for every $a,b\in\{0,1\}$.

From Eqs. (\ref{rewritten_k_gamma})-(\ref{povm_e}), the success probability of eavesdropping is obtained by
\begin{equation}
P_s^{(E)}=\frac{1-\eta_{AB}}{2(1-s^2)}(\alpha_0u_0+\alpha_1u_1).\label{soe}
\end{equation}

\subsection*{A.2. Optimization}
According to Eq. (\ref{tilde_psi}), inner product $\langle\widetilde{\psi}_0|\widetilde{\psi}_1\rangle$ is obtained by
\begin{equation}
\langle\widetilde{\psi}_0|\widetilde{\psi}_1\rangle=-s.
\end{equation}
Since $|u_0\rangle\bot|0\rangle_E$ and $|u_1\rangle\bot|0\rangle_E$, supports of $M_0^{(E)}$ and $M_1^{(E)}$ are also orthogonal to $|0\rangle_E$. This implies that Eve's POVM is designed to discriminate $|\widetilde{\psi}_0\rangle$ and $|\widetilde{\psi}_1\rangle$. Therefore, the constraint of Eve's POVM is given by  \cite{m.namkung}
\begin{eqnarray}
(1-u_0)(1-u_1)\ge s^2.\label{surface}
\end{eqnarray}
Therefore, we obtain the following optimization problem:
\begin{eqnarray}
\mathrm{maximize }&& \ \  P_s^{(E)}=\frac{1-\eta_{AB}}{2(1-s^2)}(\alpha_0u_0+\alpha_1u_1),\nonumber\\
\mathrm{subject \ to }&&\ \  (1-u_0)(1-u_1)\ge s^2.
\end{eqnarray}
For fixed parameters $\alpha_0$ and $\alpha_1$, an optimal point $(u_0,u_1)$ satisfies
\begin{equation}
(1-u_0)(1-u_1)=s^2.\label{u0u1}
\end{equation}
Also, for the optimal point, there exists a non-zero real number $\lambda$ satisfying
\begin{equation}
\vec{\nabla}P_s^{(E)}=\lambda\vec{\nabla}\left\{(1-u_0)(1-u_1)-s^2\right\},\label{gradient}
\end{equation} 
where $\vec{\nabla}$ is a gradient such that $\vec{\nabla}f=\left(\frac{\partial f}{\partial u_0},\frac{\partial f}{\partial u_1}\right)$. We note that Eq. (\ref{gradient}) is equivalent to  \cite{m.namkung}
\begin{equation}
\frac{\partial P_s^{(E)}/\partial u_0}{\partial P_s^{(E)}/\partial u_1}=\frac{\partial\left\{(1-u_0)(1-u_1)-s^2\right\}/\partial u_0}{\partial\left\{(1-u_0)(1-u_1)-s^2\right\}/\partial u_1}.\label{gradient_eq}
\end{equation}
Combining Eq. (\ref{u0u1}) and Eq. (\ref{gradient_eq}), we obtain the optimal point by
\begin{equation}
u_0=1-\sqrt{\frac{\alpha_1}{\alpha_0}}s, \ \ u_1=1-\sqrt{\frac{\alpha_0}{\alpha_1}}s.\label{opt_u0u1}
\end{equation}
Since the optimal point $(u_0,u_1)$ is on the surface of Eq. (\ref{surface}), both $u_0$ and $u_1$ in Eq. (\ref{opt_u0u1}) should be non-negative. For this reason, the overlap $s$ also should be
\begin{equation}
s<\sqrt{\frac{\alpha_0}{\alpha_1}} \ \land \ s<\sqrt{\frac{\alpha_1}{\alpha_0}}.\label{range_s}
\end{equation}
Considering $s$ in the region of Eq. (\ref{range_s}), the optimal success probability of eavesdropping is analytically given by
\begin{equation}
P_{s,opt1}^{(E)}=\frac{1-\eta_{AB}}{2(1-s^2)}(\alpha_0+\alpha_1-2\sqrt{\alpha_0\alpha_1}s).\label{p_opt1}
\end{equation}
Suppose that Bob performs optimal unambiguous discrimination between two pure states $|\psi_0\rangle$ and $|\psi_1\rangle$. Then, $\alpha_0$ and $\alpha_1$ are given by  \cite{g.jaeger}
\begin{equation}
\alpha_0=1-\sqrt{\frac{q_1}{q_0}}s, \ \ \alpha_1=1-\sqrt{\frac{q_0}{q_1}}s,\label{alpha0alpha1}
\end{equation}
if
\begin{equation}
s<\sqrt{\frac{q_1}{q_0}} \ \land \ s<\sqrt{\frac{q_0}{q_1}}.
\end{equation}
Substituting Eq. (\ref{alpha0alpha1}) with $\alpha_0$ and $\alpha_1$ in Eq. (\ref{range_s}), we obtain
\begin{eqnarray}
&&f_0(s):=q_1s^3-\sqrt{q_0q_1}s^2-q_0s+\sqrt{q_0q_1}>0,\nonumber\\
&&f_1(s):=q_0s^3-\sqrt{q_0q_1}s^2-q_1s+\sqrt{q_0q_1}>0.\label{ff}
\end{eqnarray} 
We note that if one of inequalities in Eq. (\ref{ff}) does not hold, then the optimal point $(u_0,u_1)$ is given by
\begin{eqnarray}
(u_0,u_1)\in\{(1,0),(0,1)\}.
\end{eqnarray}
Substituting this optimal point into Eq. (\ref{soe}), we obtain the optimal success probability of eavesdropping:
\begin{eqnarray}
P_{s,opt2}^{(E)}=\frac{1-\eta_{AB}}{2}\max\{\alpha_0,\alpha_1\}.\label{p_opt2}
\end{eqnarray}

\subsection*{A.3. Describing type-II structure of eavesdropper's scheme}
In this structure, the following bipartite state between Bob and Eve is considered:
\begin{equation}
\sigma_{a,BE}=\eta_{AB}|\psi_a\rangle\langle\psi_a|_B\otimes|0\rangle\langle 0|_E+(1-\eta_{AB})|\phi_+\rangle\langle\phi_+|_{BE}.
\end{equation}
Then, each $K_b^{(B)}\otimes\mathbb{I}_E\sigma_{a,AB}K_b^{(B)\dagger}\otimes\mathbb{I}_E$ is obtained by
\begin{eqnarray}
K_0\otimes\mathbb{I}_E\sigma_{0,BE}K_0^{\dagger}\otimes\mathbb{I}_E&=&\eta_{AB}\alpha_0|\phi_0^{(B)}\rangle\langle\phi_0^{(B)}|\otimes|0\rangle\langle0|_E\nonumber\\&+&\frac{(1-\eta_{AB})\alpha_0}{2(1-s^2)}|\phi_0^{(B)}\rangle\langle\phi_0^{(B)}|\otimes|\widetilde{\psi}_0\rangle\langle\widetilde{\psi}_0|_E,\nonumber\\
K_1\otimes\mathbb{I}_E\sigma_{0,BE}K_1^{\dagger}\otimes\mathbb{I}_E&=&\frac{(1-\eta_{AB})\alpha_1}{2(1-s^2)}|\phi_1^{(B)}\rangle\langle\phi_1^{(B)}|\otimes|\widetilde{\psi}_1\rangle\langle\widetilde{\psi}_1|_E,\nonumber\\
K_0\otimes\mathbb{I}_E\sigma_{1,BE}K_0^{\dagger}\otimes\mathbb{I}_E
&=&\frac{(1-\eta_{AB})\alpha_0}{2(1-s^2)}|\phi_0^{(B)}\rangle\langle\phi_0^{(B)}|\otimes|\widetilde{\psi}_0\rangle\langle\widetilde{\psi}_0|_E,\nonumber\\
K_1\otimes\mathbb{I}_E\sigma_{1,BE}K_1^{\dagger}\otimes\mathbb{I}_E
&=&\eta_{AB}\alpha_1|\phi_1^{(B)}\rangle\langle\phi_1^{(B)}|\otimes|0\rangle\langle0|_E\nonumber\\&+&\frac{(1-\eta_{AB})\alpha_1}{2(1-s^2)}|\phi_1^{(B)}\rangle\langle\phi_1^{(B)}|\otimes|\widetilde{\psi}_1\rangle\langle\widetilde{\psi}_1|_E.\label{k_sigma}
\end{eqnarray}
We derive the success probability of eavesdropping as
\begin{eqnarray}
P_s^{(E)}=\sum_{a,b\in\{0,1\}}q_a\mathrm{tr}\left[K_b^{(B)}\otimes\mathbb{I}_E\sigma_{a,BE}K_b^{(B)\dagger}\otimes\mathbb{I}_E\right]\mathrm{tr}\left[\tau_{ab,E}M_b^{(E)}\right],\label{soe_mix}
\end{eqnarray}
where $\tau_{ab,E}$ are defined as
\begin{eqnarray}
&&\tau_{00,E}=\frac{\eta_{AB}}{\eta_{AB}+\frac{1-\eta_{AB}}{2(1-s^2)}}|0\rangle\langle0|_E+\frac{\frac{1-\eta_{AB}}{2(1-s^2)}}{\eta_{AB}+\frac{1-\eta_{AB}}{2(1-s^2)}}|\widetilde{\psi}_0\rangle\langle\widetilde{\psi}_0|_E,\nonumber\\
&&\tau_{01,E}=|\widetilde{\psi}_1\rangle\langle\widetilde{\psi}_1|_E,\nonumber\\
&&\tau_{10,E}=|\widetilde{\psi}_0\rangle\langle\widetilde{\psi}_0|_E,\nonumber\\
&&\tau_{11,E}=\frac{\eta_{AB}}{\eta_{AB}+\frac{1-\eta_{AB}}{2(1-s^2)}}|0\rangle\langle0|_E+\frac{\frac{1-\eta_{AB}}{2(1-s^2)}}{\eta_{AB}+\frac{1-\eta_{AB}}{2(1-s^2)}}|\widetilde{\psi}_1\rangle\langle\widetilde{\psi}_1|_E.\label{tautau}
\end{eqnarray}
From Eqs. (\ref{k_sigma}) and (\ref{tautau}), the success probability of eavesdropping in Eq. (\ref{soe_mix}) is obtained by Eq. (\ref{soe}).

\section*{Appendix B. Secret key rate}
In this section, we derive the secret key rate when Eve's POVM optimizes the success probability of eavesdropping.

\subsection*{B.1. Secret key rate of type-I eavesdropping structure}
To derive the secret key rate, we need to evaluate entropies $H(A)$, $H(B,A)$, $H(E)$ and $H(B,E)$. For equal prior probabilities (\textit{i.e}., $q_0=q_1$), $H(A)$ is given by
\begin{equation}
H(A)=-q_0\log_2q_0-q_1\log_2q_1=1.\label{ha}
\end{equation}

Also, $H(B,A)$ is given by
\begin{equation}
H(B,A)=-\sum_{a,b\in\{0,1\}}\widetilde{P}_{AB}(a,b)\log_2\widetilde{P}_{AB}(a,b),\label{hab}
\end{equation}
where $\widetilde{P}_{AB}(a,b)$ is a post-processed joint probability between Alice and Bob after Bob discards his inconclusive result:
\begin{equation}
\widetilde{P}_{AB}(a,b)=\frac{P_{AB}(a,b)}{\sum_{a,b\in\{0,1\}}P_{AB}(a,b)},
\end{equation}
and $P_{AB}(a,b)$ is a pre-processed joint probability
\begin{eqnarray}
P_{AB}(a,b)=q_a\mathrm{tr}\left\{\Lambda^{(A\rightarrow B)}(|\psi_a\rangle\langle\psi_a|)K_b^{(B)\dagger}K_b^{(B)}\right\}=\frac{1}{2}\left\{\eta_{AB}\alpha_b\delta_{ab}+\frac{(1-\eta_{AB})\alpha_b}{2(1-s^2)}\right\}.\label{p_ab}
\end{eqnarray}
Here, we consider the post-processing that Bob discard his inconclusive result, since this post-processing can enhance unambiguous quantum communication protocol  \cite{d.fields}.

Since $q_0=q_1$ implies $u_0=u_1=1-s$ according to Eq. (\ref{opt_u0u1}) and Eq. (\ref{alpha0alpha1}), the joint probability of Eq. (\ref{p_ab}) is rewritten by
\begin{eqnarray}
P_{AB}(a,b)&=& 
      \frac{1}{2}\left\{\eta_{AB}(1-s)+\frac{1-\eta_{AB}}{2(1+s)}\right\}, \ \  \mathrm{if} \ \  (a,b)\in\{(0,0),(1,1)\},
      \\
      P_{AB}(a,b)&=&\frac{1-\eta_{AB}}{4(1+s)}, \ \  \mathrm{if} \ \  (a,b)\in\{(0,1),(1,0)\}.
\label{19}
\end{eqnarray}

To evaluate $H(B,E)$ and $H(E)$, we first consider a joint probability $P_{ABE}(a,b,e)$ among Alice, Bob and Eve:
\begin{eqnarray}
P_{ABE}(a,b,e)=P_{AB}(a,b)P_{E|AB}(e|a,b)=q_aP_{B|A}(b|a)P_{E|AB}(e|a,b),\label{p_abe}
\end{eqnarray}
where $P_{B|A}(b|a)$ and $P_{E|AB}(e|a,b)$ are conditional probabilities. 

\begin{enumerate}
\item In case of $b\not=?$, every $|\gamma_{ab}\rangle$ in Eq. (\ref{gamma_vector}) is rewritten by
\begin{equation}
|\gamma_{ab}\rangle_E=\frac{1}{\sqrt{P_{B|A}(b|a)}}\left\{\sqrt{\eta_{AB}\alpha_b}\delta_{ab}|0\rangle+\sqrt{\frac{(1-\eta_{AB})\alpha_b}{2(1-s^2)}}|\widetilde{\psi}_b\rangle\right\}_E,
\end{equation}
where
\begin{equation}
P_{B|A}(b|a)=\eta_{AB}\alpha_b\delta_{ab}+\frac{(1-\eta_{AB})\alpha_b}{2(1-s^2)}.\label{pb|a}
\end{equation}
Therefore, $P_{E|AB}(e|a,b)$ is given by
\begin{equation}
P_{E|AB}(e|a,b)=\langle\gamma_{ab}|M_e^{(E)}|\gamma_{ab}\rangle=\frac{1}{P_{B|A}(b|a)}\frac{(1-\eta_{AB})\alpha_b}{2(1-s^2)}u_e\delta_{be}.\label{e_ab}
\end{equation}
Substituting Eq. (\ref{e_ab}) into $P_{E|AB}(e|a,b)$ of Eq. (\ref{p_abe}), we obtain 
\begin{equation}
P_{ABE}(a,b,e)=q_a\frac{(1-\eta_{AB})\alpha_b}{2(1-s^2)}u_e\delta_{be},
\end{equation}
and
\begin{equation}
P_{BE}(b,e)=\sum_{a=0}^{1}P_{ABE}(a,b,e)=\frac{(1-\eta_{AB})\alpha_b}{2(1-s^2)}u_e\delta_{be},\label{p_be1}
\end{equation}
\item in case of $b=?$, we provide following equality:
\begin{eqnarray}
&&K_?^{(B)}\otimes\mathbb{I}_E|\Gamma_a\rangle\nonumber\\
&&=\sqrt{\eta_{AB}(1-\alpha_0)}\delta_{a0}|\phi_0^{(B)}\rangle_B\otimes|0\rangle_E+\sqrt{\frac{(1-\eta_{AB})(1-\alpha_0)}{2(1-s^2)}}|\phi_0^{(B)}\rangle_B\otimes|\widetilde{\psi}_0\rangle_E,\nonumber\\
&&+\sqrt{\eta_{AB}(1-\alpha_1)}\delta_{a1}|\phi_1^{(B)}\rangle_B\otimes|0\rangle_E+\sqrt{\frac{(1-\eta_{AB})(1-\alpha_1)}{2(1-s^2)}}|\phi_1^{(B)}\rangle_B\otimes|\widetilde{\psi}_1\rangle_E.\nonumber\\
\end{eqnarray}
In the same way as Eq. (\ref{gamma_vector}), we obtain
\begin{eqnarray}
|\gamma_{a?}\rangle_E=\frac{1}{P_{B|A}(?|a)}\sum_{x\in\{0,1\}}\left\{\sqrt{\eta_{AB}(1-\alpha_x)}\delta_{ax}|0\rangle+\sqrt{\frac{(1-\eta_{AB})(1-\alpha_x)}{2(1-s^2)}}|\widetilde{\psi}_x\rangle\right\}_E.
\end{eqnarray}
(Since $P_{B|A}(?|a)$ is too complicated, we do not describe it in detail.) Therefore, $P_{E|AB}(e|a,?)$ is given by
\begin{eqnarray}
P_{E|AB}(e|a,?)=\frac{1}{P_{B|A}(?|a)}\left\{\frac{(1-\eta_{AB})(1-\alpha_0)}{2(1-s^2)}u_e\delta_{e0}+\frac{(1-\eta_{AB})(1-\alpha_1)}{2(1-s^2)}u_e\delta_{e1}\right\}.\label{e_a?}
\end{eqnarray}
Substituting Eq. (\ref{e_a?}) into $P_{E|AB}(e|a,b)$ of Eq. (\ref{p_abe}), we obtain 
\begin{eqnarray}
P_{ABE}(a,b,e)=q_a\left\{\frac{(1-\eta_{AB})(1-\alpha_0)}{2(1-s^2)}u_e\delta_{0e}+\frac{(1-\eta_{AB})(1-\alpha_1)}{2(1-s^2)}u_e\delta_{1e}\right\},
\end{eqnarray}
and
\begin{eqnarray}
P_{BE}(?,e)&=&\sum_{a\in\{0,1\}}P_{ABE}(a,?,e)\nonumber\\
&=&\frac{(1-\eta_{AB})(1-\alpha_0)}{2(1-s^2)}u_e\delta_{0e}+\frac{(1-\eta_{AB})(1-\alpha_1)}{2(1-s^2)}u_e\delta_{1e},\label{p_be2}
\end{eqnarray}
\end{enumerate}
Since $q_0=q_1$ implies $u_0=u_1=1-s$ according to Eq. (\ref{opt_u0u1}) and Eq. (\ref{alpha0alpha1}), the joint probability $P_{BE}(b,e)$ of Eq. (\ref{p_be1}) and Eq. (\ref{p_be2}) are rewritten by
\begin{eqnarray}
P_{BE}(b,e)&=&
      \frac{(1-\eta_{AB})(1-s)}{2(1+s)}, \ \  \mathrm{if} \ \  (b,e)\in\{(0,0),(1,1)\},\nonumber
      \\
     P_{BE}(b,e)&=&  0, \ \  \mathrm{if}\enspace (b,e)\in\{(0,1),(1,0)\},\nonumber
      \\
     P_{BE}(b,e)&=&  \frac{(1-\eta_{AB})s}{2(1+s)}, \ \  \mathrm{if} \ \  (b,e)\in\{(0,?),(1,?)\}.
\label{190}
\end{eqnarray}
If Eve discard her inconclusive result, the post-processed joint probability is given by
\begin{equation}
\widetilde{P}_{BE}(b,e)=\frac{P_{BE}(b,e)}{\sum_{b\in\{0,1,?\}}\sum_{e\in\{0,1\}}P_{BE}(b,e)},
\end{equation}
and a marginal probability $\widetilde{P}_E(e)$ is given by
\begin{equation}
\widetilde{P}_E(e)=\sum_{b\in\{0,1,?\}}\widetilde{P}_{BE}(b,e).
\end{equation}
Finally, $H(E)$ and $H(B,E)$ are evaluated as
\begin{eqnarray}
H(E)&=&-\sum_{e\in\{0,1\}}\widetilde{P}_{E}(e)\log_2\widetilde{P}_{E}(e),\nonumber\\
H(B,E)&=&-\sum_{b\in\{0,1,?\}}\sum_{e\in\{0,1\}}\widetilde{P}_{BE}(b,e)\log_2\widetilde{P}_{BE}(b,e).
\end{eqnarray}

\subsection*{B.2. Secret key rate of type-II eavesdropping structure}
We first note that the prior probabilities and the quantum channel $\Lambda^{(A\rightarrow B)}$ are invariant under the choice of structure. Therefore, $H(A)$ and $H(B,A)$ are evaluated as Eq. (\ref{ha}) and Eq. (\ref{hab}) in the type-I structure.

\begin{enumerate}
\item In case of $b\not=?$, every $\tau_{ab,E}$ in Eq. (\ref{tautau}) is rewritten by
\begin{equation}
\tau_{ab,E}=\frac{1}{P_{B|A}(b|a)}\left\{\eta_{AB}\alpha_b\delta_{ab}|0\rangle\langle 0|+\frac{(1-\eta_{AB})\alpha_b}{2(1-s^2)}|\widetilde{\psi}_b\rangle\langle\widetilde{\psi}_b|\right\},
\end{equation}
where $P_{B|A}(b|a)$ is given by Eq. (\ref{pb|a}). Moreover, $P_{E|AB}(e|a,b)$ is given by
\begin{equation}
P_{E|AB}(e|a,b)=\mathrm{tr}\left\{\tau_{ab,E}M_e^{(E)}\right\}=\frac{1}{P_{B|A}(b|a)}\frac{(1-\eta_{AB})\alpha_b}{2(1-s^2)}u_e\delta_{be},\label{e_ab_tau}
\end{equation}
which is equal to Eq. (\ref{e_ab}). Therefore, according to Eq. (\ref{p_abe}), \textit{$P_{ABE}(a,b,e)$ is equal to the case of type-I structure.}
\item In case of $b=?$, we consider
\begin{eqnarray}
&&K_?^{(B)}\otimes\mathbb{I}_E\sigma_{a,BE}K_?^{(B)}\otimes\mathbb{I}_E\nonumber\\&&=\eta_{AB}\Gamma(\sqrt{1-\alpha_0}\delta_{a0}|\phi_0^{(B)}\rangle+\sqrt{1-\alpha_1}\delta_{a1}|\phi_1^{(B)}\rangle)\otimes|0\rangle\langle 0|_E\nonumber\\
&&+(1-\eta_{AB})\Gamma\left(\sqrt{\frac{1-\alpha_0}{2(1-s^2)}}|\phi_0^{(B)}\rangle\otimes|\widetilde{\psi}_0\rangle+\sqrt{\frac{1-\alpha_1}{2(1-s^2)}}|\phi_1^{(B)}\rangle\otimes|\widetilde{\psi}_1\rangle\right),
\end{eqnarray}
where we define $\Gamma(|v\rangle):=|v\rangle\langle v|$ for convenience. From the above representation, we define a bipartite mixed state shared by Bob and Eve:
\begin{eqnarray}
\tau_{a?,BE}&=&\frac{1}{P_{B|A}(b|a)}\Big[\eta_{AB}\Gamma(\sqrt{1-\alpha_0}\delta_{a0}|\phi_0^{(B)}\rangle+\sqrt{1-\alpha_1}\delta_{a1}|\phi_1^{(B)}\rangle)\otimes|0\rangle\langle 0|_E\nonumber\\
&+&(1-\eta_{AB})\Gamma\left(\sqrt{\frac{1-\alpha_0}{2(1-s^2)}}|\phi_0^{(B)}\rangle\otimes|\widetilde{\psi}_0\rangle+\sqrt{\frac{1-\alpha_1}{2(1-s^2)}}|\phi_1^{(B)}\rangle\otimes|\widetilde{\psi}_1\rangle\right)\Big].\nonumber\\
\end{eqnarray}
Then, $P_{E|AB}(e|a,?)$ is given by
\begin{eqnarray}
P_{E|AB}(e|a,?)&=&\mathrm{tr}\left\{\tau_{a?,BE}\left(\mathbb{I}_B\otimes M_e^{(E)}\right)\right\}\nonumber\\
&=&\frac{1}{P_{B|A}(?|a)}\left\{\frac{(1-\eta_{AB})(1-\alpha_0)}{2(1-s^2)}u_e\delta_{e0}+\frac{(1-\eta_{AB})(1-\alpha_1)}{2(1-s^2)}u_e\delta_{e1}\right\}.\nonumber\\
\end{eqnarray}
This is equal to Eq. (\ref{e_a?}), which implies that $P_{A,B,E}(a,b,?)$ \textit{is also equal to the case of type-I structure.}
\end{enumerate}

From the above calculation, we confirm that $H(B,E)$ and $H(E)$ in this structure is equal to these in the type-I structure, respectively. This leads us to the result that both structures provide same secret key rate.

It is noted that the formalism of the joint probabilities discussed above can also provide the success probability of eavesdropping. This will be further explained in the next section.

\subsection*{B.3. Revisiting success probability of eavesdropping in terms of joint probabilities}
In the scenario of the new sequential discrimination, Bob's measurement result $b$ depends on the input $a$ prepared by Alice, and Eve's measurement result $e$ depends on $a$ and $b$. From these facts, the joint probability between three parties $P_{ABE}(a,b,e)$ is easily derived by
\begin{equation}
P_{ABE}(a,b,e)=q_aP_{B|A}(b|a)P_{E|AB}(e|a,b)=q_aP_{BE|A}(b,e|a),
\end{equation}
where $q_a$ is the prior probability that Alice prepares $a$, $P_{B|A}(b|a)$ is the conditional probability that Bob obtains $b$ if Alice prepares $a$, $P_{E|AB(e|a,b}$ is the conditional probability that Eve obtains $e$ if Alice prepares $a$ and Bob obtains $b$, $P_{BE|A}(b,e|a$ is the conditional joint probability that Bob and Eve obtain $b$ and $e$ if Alice prepares $a$, and $P_{BE}(b,e)$ is the joint probability that Bob and Eve obtain $b$ and $e$. Also, the success probability of eavesdropping is derived by
\begin{eqnarray}
P_s^{(E)}=\sum_{a,b}q_aP_{B|A}(b|a)P_{E|AB}(e=b|a,b)=\sum_{a,b}q_aP_{BE|A}(b,e=b|a)=\sum_{b}P_{BE}(b,e=b).\label{gen}
\end{eqnarray}
It is noted that the expression of the success probability of eavesdropping in Eq. (\ref{gen}) is used for deriving the success probability of eavesdropping when Alice, Bob, and Eve performs the scenario by using the imperfect linear optical technologies.

\section*{Appendix C. Description of imperfect quantum optical setting}

\subsection*{C.1. Elementary linear optical devices}
First, we provides some conventions of linear optical devices including a half wave plate, a polarized beam splitter and a single photon detector. 
\begin{figure}[t]
\includegraphics[width=15cm]{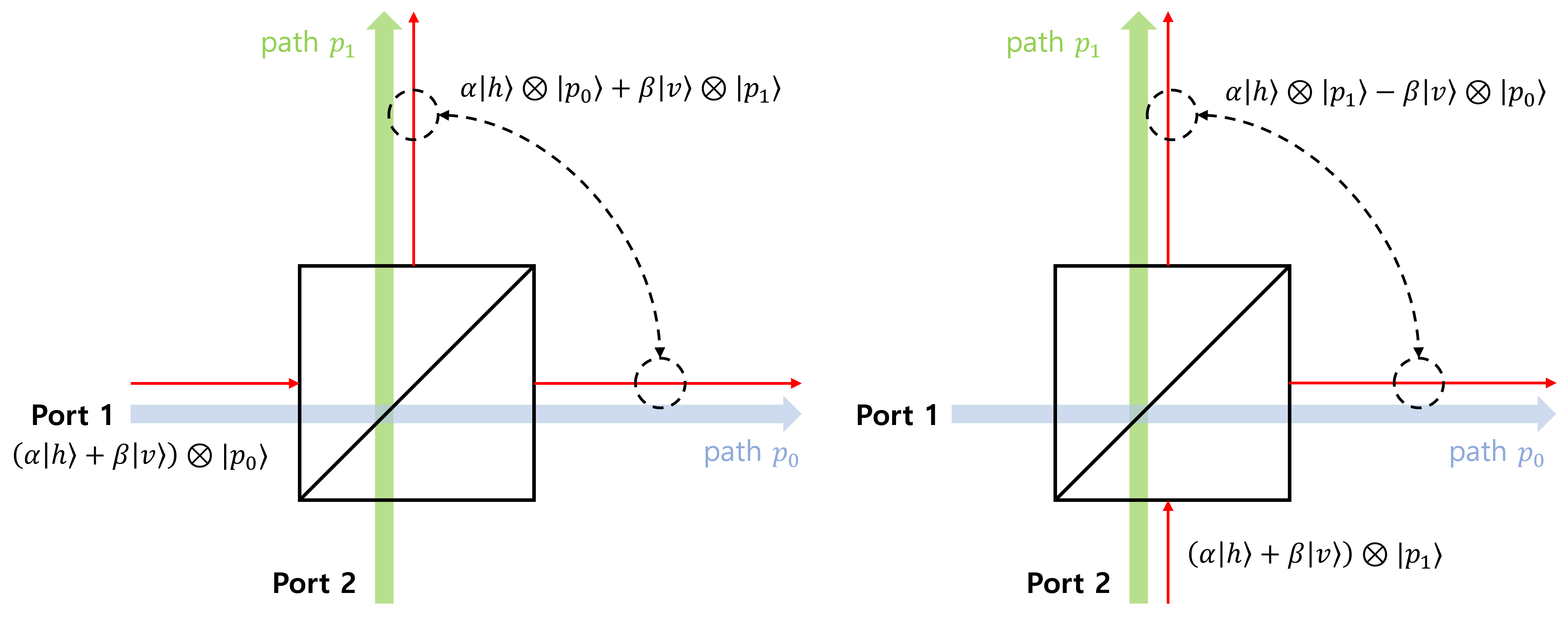}
\caption{Illustration of a polarized beam splitter.}
\centering
\end{figure}
\begin{enumerate}
\item Let that horizontally and vertically polarized single photon states be denoted as $|h\rangle$ and $|v\rangle$, respectively. Then, a half wave plate transforms these states to
\begin{equation}
|h\rangle\rightarrow\cos2\gamma|h\rangle+\sin2\gamma|v\rangle, \ \ |v\rangle\rightarrow\sin2\gamma|h\rangle-\cos2\gamma|v\rangle,
\end{equation}
where $\gamma$ is a polarization angle of the half wave plate  \cite{w.-h.zhang}.
\item We use the convention of a polarized beam splitter as illustrated in Fig. 8. We note that the phase of the vertically polarized state changes as $|v\rangle\rightarrow-|v\rangle$ if the single photon enters into the port 2 of the polarized beam splitter. This happens because of the energy conservation law  \cite{a.furusawa}: the entire energy of input and output lights should be equal.
\item By using POVM representation, we generally describe the imperfect on/off detector $\{\Pi_{\rm off},\Pi_{\rm on}\}$  \cite{g.cariolaro}:
\begin{eqnarray}
\Pi_{\rm off}(\nu,\eta)=e^{-\nu}\sum_{n=0}^{\infty}(1-\eta)^n|n\rangle\langle n|, \ \ \Pi_{\rm on}(\nu,\eta)=\mathbb{I}-\Pi_{\rm off}(\nu,\eta),\label{onoff}
\end{eqnarray} 
where $\{|n\rangle|n=0,1,2,\cdots\}$ is Fock basis and $\mathbb{I}$ is an identity operator composed of the Fock basis. In the representation, $\nu\ge0$ and $\eta\in[0,1]$ are dark count rate and detection efficiency, respectively. If $\nu=0$ and $\eta=1$,  Eq. (\ref{onoff}) represents an ideal on/off detector without dark count and probabilistic non-detection: 
\begin{eqnarray}
\Pi_{\rm off}(0,1)=|0\rangle\langle0|, \ \ \Pi_{\rm on}(0,1)=\mathbb{I}-|0\rangle\langle0|.
\end{eqnarray}

Here, we consider $\nu\simeq0$, since current quantum technologies can provide us to implement a superconducting nanowire single photon detector with very low chance of dark count  \cite{snspd}. Alice encodes her message in a single photon, and thus Bob and Eve's detectors need to distinguish whether there is a photon in a path or not. For this reason, each detector of represented in Eq. (\ref{onoff}) is rewritten by
\begin{equation}
\Pi_{\rm off}(\eta):=|0\rangle\langle0|+(1-\eta)|1\rangle\langle1|, \ \ \Pi_{\rm on}(\eta):=\eta|1\rangle\langle 1|.
\end{equation}
\end{enumerate}

\subsection*{C.2. Sagnac-like interferometers of Bob and Eve}
In the entire optical setting, Bob's Sagnac-like interemetor performs the transformation
\begin{eqnarray}
|h\rangle\otimes|b_0\rangle&\rightarrow&\sqrt{\frac{1-s}{1+s}}|h\rangle\otimes|b_0\rangle+\sqrt{\frac{2s}{1+s}}|v\rangle\otimes|b_1\rangle,\nonumber\\
|v\rangle\otimes|b_0\rangle&\rightarrow&|v\rangle\otimes|b_0\rangle,\label{sag_b}
\end{eqnarray} 
and Eve's Sagnac-like interferometer performs the transformation
\begin{eqnarray}
|h\rangle\otimes|e_0\rangle&\rightarrow&|h\rangle\otimes|e_0\rangle,\nonumber\\
|v\rangle\otimes|e_0\rangle&\rightarrow&\sqrt{\frac{2s}{1+s}}|h\rangle\otimes|e_1\rangle+\sqrt{\frac{1-s}{1+s}}|v\rangle\otimes|e_0\rangle,\label{sag_e}
\end{eqnarray}
where $b_0$ ($e_0$) denotes the clockwise path and $b_1$ ($e_1$) denotes the counter-clockwise in Bob's (Eve's) Sagnac-like interferometer. 

\subsection*{C.3. Noisy entanglement between Bob and Eve}
We also consider that an imperfect entangled state takes the form:
\begin{equation}
\varrho_{ent}^{(wh)}=\eta_{\rm ent} |\phi_+\rangle\langle\phi_+|+(1-\eta_{\rm ent})\frac{1}{4}(|h\rangle\langle h|+|v\rangle\langle v|)\otimes(|h\rangle\langle h|+|v\rangle\langle v|),\label{wh}
\end{equation}
or
\begin{equation}
\varrho_{ent}^{(cl)}=\eta_{\rm ent} |\phi_+\rangle\langle\phi_+|+(1-\eta_{\rm ent})\frac{1}{2}(|hh\rangle\langle hh|+|vv\rangle\langle vv|),\label{cl}
\end{equation}
where $|\phi_+\rangle\propto|hh\rangle+|vv\rangle$ and $\eta_{\rm ent}$ is a magnitude of decoherence. The noise described in Eq. (\ref{wh}) and Eq. (\ref{cl}) are called white noise and color noise  \cite{a.cabello}, respectively. 

\section*{Appendix D. Success probability of eavesdropping and secret key rate in the imperfect optical setting}
We only provide a methodoloy for evaluating the success probability and the secret key rate in the imperfect optical setting, since entire derivation is too lengthy and complicated. Suppose that Eve intercepts Alice's state with a probability $1-\eta_{AB}$ and shares her noisy entangled state in Eq. (\ref{wh}) or Eq. (\ref{cl}) with Bob. Then, Bob and Eve eventually shares the following ensemble:
\begin{equation}
\zeta_{\pm}:=\eta_{AB}\Lambda_{D_0}^{(ad)}(|\psi_\pm\rangle\langle\psi_\pm|)_B\otimes|0\rangle\langle 0|_E+(1-\eta_{AB})\left(\Lambda_{D_e}^{(ad)}\otimes\Lambda_{D_e}^{(ad)}\right)(\rho_{ent}),
\end{equation}
where $\Lambda_{D}^{(ad)}$ is the amplitude damping channel with decoherence rate $D\in[0,1]$  \cite{y.-s.kim}. Here, large $D$ means that the strength of the amplitude damping is large. Let us denote the initial path state of Bob as $|b_0\rangle\langle b_0|_{P_B}$ where $P_B$ is a path system of Bob. If Bob obtains a conclusive result, then $\zeta_{\pm}\otimes|b_0\rangle\langle b_0|$ is transformed to a sub-normalized positive-semidefinite operator
\begin{eqnarray}
\xi_{con,\pm}\label{xi_c}=\mathrm{Tr}_{P_B}\left\{\mathbb{I}_{BE}\otimes|b_0\rangle\langle b_0|_{P_B}\left(U^{(sag)}\otimes\mathbb{I}_{EP_B}\right)\zeta_{\pm}\otimes|b_0\rangle\langle b_0|\left(U^{(sag)}\otimes\mathbb{I}_{EP_B}\right)^\dagger\right\},
\end{eqnarray}
and if Bob obtains an inconclusive result, then $\zeta_{\pm}\otimes|b_0\rangle\langle b_0|$ is transformed to 
\begin{eqnarray}
\xi_{inc,\pm}\label{xi_i}=\mathrm{Tr}_{P_B}\left\{\mathbb{I}_{BE}\otimes|b_1\rangle\langle b_1|_{P_B}\left(U^{(sag)}\otimes\mathbb{I}_{EP_B}\right)\zeta_{\pm}\otimes|b_0\rangle\langle b_0|\left(U^{(sag)}\otimes\mathbb{I}_{EP_B}\right)^\dagger\right\}.
\end{eqnarray}
In Eqs. (\ref{xi_c}) and (\ref{xi_i}), $\mathbb{I}_{XY}$ is an identity operator on the composite system $XY$ with $X,Y\in\{B,E,P_B\}$ and $U^{(sag)}$ is the unitary operator of the Sagnac-like interferometer performing the transformation in Eq. (\ref{sag_b}). Bob's half wave plate and polarized beam splitter transforms $\xi_{con,\pm}$ and $\xi_{inc,\pm}$ to
\begin{eqnarray}
\xi_{con,\pm}&\rightarrow\Omega_{con,\pm}:= \left(V^{(PBS)}H^{(HWP)}\otimes\mathbb{I}_E\right)\xi_{con,\pm}\left(V^{(PBS)}H^{(HWP)}\otimes\mathbb{I}_E\right)^\dagger,\nonumber\\
\xi_{inc,\pm}&\rightarrow\Omega_{inc,\pm}:= \left(V^{(PBS)}H^{(HWP)}\otimes\mathbb{I}_E\right)\xi_{inc,\pm}\left(V^{(PBS)}H^{(HWP)}\otimes\mathbb{I}_E\right)^\dagger,
\label{trans}
\end{eqnarray}
where $H^{(HWP)}$ is the Hadamard operator describing the half wave plate and $V^{(PBS)}$ is the isometry:
\begin{eqnarray}
V^{(PBS)}&:&|h\rangle\rightarrow|1,0\rangle_{B_0B_1}=|1\rangle_{B_0}\otimes|0\rangle_{B_1},\nonumber\\
V^{(PBS)}&:&|v\rangle\rightarrow-|0,1\rangle_{B_0B_1}=-|0\rangle_{B_0}\otimes|1\rangle_{B_1}.\label{isom}
\end{eqnarray}
In the right hand sides, $B_0$ and $B_1$ denotes the two output ports of the Bob's PBS (for detail, see Fig. 8), and $|n\rangle$ ($n=0,1$) denotes the photon number state. In Eq. (\ref{isom}), minus sign implies the energy conservation law in the Bob's PBS, which was already discussed in the item 2 of the Section C. From the relation in Eq. (\ref{trans}), the conditional probability that bob obtains a conclusive result $b\in\{0,1\}$ if Alice prepares $a\in\{0,1\}$ is evaluated as
\begin{eqnarray}
&&P_{B|A}(b=0|a)=\mathrm{Tr}_{B_0B_1}\left\{\Pi_{\rm on}(\eta)\otimes\Pi_{\rm off}(\eta)\otimes\mathbb{I}_E\Omega_{con,\pm}\right\},\nonumber\\
&&P_{B|A}(b=1|a)=\mathrm{Tr}_{B_0B_1}\left\{\Pi_{\rm off}(\eta)\otimes\Pi_{\rm on}(\eta)\otimes\mathbb{I}_E\Omega_{con,\pm}\right\}.
\end{eqnarray}

To compute the marginal conditional probability between Eve and Bob, let us consider the sub-normalized positive-semidefinite operators:
\begin{eqnarray}
\widetilde{\zeta}_{\pm}(b=0|a)&=&\mathrm{Tr}_{B_0B_1}\left\{\Pi_{\rm on}(\eta)\otimes\Pi_{\rm off}(\eta)\otimes\mathbb{I}_E\Omega_{con,\pm}\right\},\nonumber\\
\widetilde{\zeta}_{\pm}(b=1|a)&=&\mathrm{Tr}_{B_0B_1}\left\{\Pi_{\rm off}(\eta)\otimes\Pi_{\rm on}(\eta)\otimes\mathbb{I}_E\Omega_{con,\pm}\right\},\nonumber\\
\widetilde{\zeta}_{\pm}(b=?|a)&=&\mathrm{Tr}_{B_0B_1}\left\{\xi_{inc,\pm}\right\}.
\end{eqnarray}
Here, $b=?$ denotes Bob's inconclusive result. Let us denote the initial path state of Eve as $|e_0\rangle\langle e_0|_{P_E}$ where $P_E$ is a path system of Eve. If Eve obtains a conclusive result, then $\widetilde{\zeta}_{\pm}(b|a)\otimes|e_0\rangle\langle e_0|$ is transformed to a sub-normalized positive-semidefinite operator
\begin{eqnarray}
\widetilde{\xi}_{con,\pm}(b|a)\label{con_e}=\mathrm{Tr}_{P_E}\left\{\mathbb{I}_{E}\otimes|e_0\rangle\langle e_0|_{P_E}\left(\widetilde{U}^{(sag)}\otimes\mathbb{I}_{EP_E}\right)\widetilde{\zeta}_{\pm}(b|a)\otimes|e_0\rangle\langle e_0|\left(\widetilde{U}^{(sag)}\otimes\mathbb{I}_{EP_E}\right)^\dagger\right\},
\end{eqnarray}
and if Eve obtains an inconclusive result, then $\widetilde{\zeta}_{\pm}(b|a)\otimes|e_0\rangle\langle e_0|$ is transformed to 
\begin{eqnarray}
\widetilde{\xi}_{inc,\pm}(b|a)\label{inc_e}=\mathrm{Tr}_{P_E}\left\{\mathbb{I}_{E}\otimes|e_1\rangle\langle e_1|_{P_E}\left(\widetilde{U}^{(sag)}\otimes\mathbb{I}_{EP_E}\right)\widetilde{\zeta}_{\pm}(b|a)\otimes|e_0\rangle\langle e_0|\left(\widetilde{U}^{(sag)}\otimes\mathbb{I}_{EP_E}\right)^\dagger\right\}.
\end{eqnarray}
In Eqs. (\ref{con_e}) and (\ref{inc_e}), $\widetilde{U}^{(sag)}$ is the unitary operator of the Sagnac-like interferometer performing the transformation in Eq. (\ref{sag_e}). Eve's half wave plate and polarized beam splitter transforms $\xi_{con,\pm}$ and $\xi_{inc,\pm}$ to
\begin{eqnarray}
\widetilde{\xi}_{con,\pm}(b|a)&\rightarrow&\widetilde{\Omega}_{con,\pm}:= \left(\widetilde{V}^{(PBS)}\widetilde{H}^{(HWP)}\otimes\mathbb{I}_E\right)\widetilde{\xi}_{con,\pm}(b|a)\left(\widetilde{V}^{(PBS)}\widetilde{H}^{(HWP)}\otimes\mathbb{I}_E\right)^\dagger,\nonumber\\
\widetilde{\xi}_{inc,\pm}(b|a)&\rightarrow&\widetilde{\Omega}_{inc,\pm}:=  \left(\widetilde{V}^{(PBS)}\widetilde{H}^{(HWP)}\otimes\mathbb{I}_E\right)\widetilde{\xi}_{inc,\pm}(b|a)\left(\widetilde{V}^{(PBS)}\widetilde{H}^{(HWP)}\otimes\mathbb{I}_E\right)^\dagger,
\label{trans_e}
\end{eqnarray}
where $\widetilde{H}^{(HWP)}$ is the Hadamard operator describing the half wave plate and $\widetilde{V}^{(PBS)}$ is the isometry:
\begin{eqnarray}
\widetilde{V}^{(PBS)}&:&|h\rangle\rightarrow|1,0\rangle_{E_0E_1}=|1\rangle_{E_0}\otimes|0\rangle_{E_1},\nonumber\\
\widetilde{V}^{(PBS)}&:&|v\rangle\rightarrow-|0,1\rangle_{E_0E_1}=-|0\rangle_{E_0}\otimes|1\rangle_{E_1}.\label{isom_e}
\end{eqnarray}
In the right hand sides, $E_0$ and $E_1$ denotes the two output ports of the Eve's PBS (for detail, see Fig. 8), and $|n\rangle$ ($n=0,1$) denotes the photon number state. From the relation in Eq. (\ref{trans_e}), the conditional probability that Bob and Eve obtains a conclusive result $b\in\{0,1\}$ and $e\in\{0,1\}$, respectively, if Alice prepares $a\in\{0,1\}$ is evaluated as
\begin{eqnarray}
P_{EB|A}(e=0,b|a)&=&\mathrm{Tr}_{B_0B_1}\left\{\Pi_{\rm on}(\eta)\otimes\Pi_{\rm off}(\eta)\otimes\mathbb{I}_E\rightarrow\widetilde{\Omega}_{con,\pm}\right\},\nonumber\\
P_{EB|A}(e=1,b|a)&=&\mathrm{Tr}_{B_0B_1}\left\{\Pi_{\rm off}(\eta)\otimes\Pi_{\rm on}(\eta)\otimes\mathbb{I}_E\rightarrow\widetilde{\Omega}_{con,\pm}\right\},\nonumber\\
P_{EB|A}(e=?,b|a)&=&\mathrm{Tr}_{B_0B_1}\left\{\widetilde{\xi}_{inc,\pm}(b|a)\right\}.
\end{eqnarray}

\end{document}